\documentclass[aps, prd, twocolumn, showpacs, superscriptaddress, groupedaddress]{revtex4} 
\usepackage{graphicx}    
\usepackage{amssymb}
\usepackage{subfigure}
\usepackage{float}
\usepackage[pagebackref=false, colorlinks=true]{hyperref}
\hypersetup{
linkcolor=blue,     
citecolor=blue,     
urlcolor=blue}   
\usepackage{soul}
\usepackage{amsmath}
\begin{document}
\title{Global visibility of a strong curvature singularity in  nonmarginally bound dust collapse}
\author{Karim Mosani}
\email{kmosani2014@gmail.com}
\affiliation{BITS Pilani K.K. Birla Goa Campus, Sancoale, Goa 403726, India}
\author{Dipanjan Dey}
\email{dipanjandey.icc@charusat.ac.in}
\affiliation{International Center for Cosmology, Charusat University, Anand 388421, Gujarat, India}
\author{Pankaj S. Joshi}
\email{psjprovost@charusat.ac.in}
\affiliation{International Center for Cosmology, Charusat University, Anand 388421, Gujarat, India}

\date{\today}

\begin{abstract}
We investigate here the local versus global visibility of a spacetime singularity formed due to the gravitational collapse of a spherically symmetric dust cloud having a nonzero velocity function. The conditions are investigated that ensure the global visibility of the singularity, in the sense that the outgoing null geodesics leave the boundary of the matter cloud in the future, whereas, in the past, these terminate at the singularity.  Explicit examples of this effect are constructed. We require that this must be a strong curvature singularity in the sense of Tipler, to ensure the physical significance of the scenario considered. This may act as a counterexample to the weak cosmic censorship hypothesis. 
\\
\bigskip

$\textbf{key words}$: Cosmic Censorship Hypothesis, Gravitational Collapse, Naked Singularity, Strong Singularity.

\end{abstract}
\maketitle

\section{Introduction}
When a sufficiently massive cloud collapses unhindered under the influence of its gravitational field, a spacetime singularity is obtained. The visibility of such singularity has created much interest in recent years, as that would violate the cosmic censorship hypothesis. The first model of the gravitational collapse was studied by Oppenheimer and Snyder  
\cite{oppenheimer} 
in 1939, and by Datt 
\cite{datt} 
independently. The model had zero pressure, and the density distribution was homogeneous. The end state of such a collapse turns out to be a singularity from where no nonspacelike geodesic can escape. It was argued then  that singularities were merely an artifact of the exact symmetries
\cite{hawking},
e.g., in the case of gravitational collapse, it is the assumption of spherical symmetry that might cause the occurrence of singularities. However, in a real scenario, this would not be the case. Hence it was argued that no such singularities arise in reality. However, the cosmological evidence provided by the WMAP, COBE, and Planck of the cosmic microwave background radiation indicates that the precursor of our present universe is a singularity, i.e., the universe had a singular initiation. Apart from this, the singularity theorems provided by Penrose and Hawking 
\cite{hawking, penrose} 
prove that singularities could indeed form under very generic conditions in gravitational collapse as well as cosmology. These observations, together with the above-mentioned theorems, suggest the existence of singularities in the universe. It is worth noting that the singularity theorem can be interpreted in two different ways. One could interpret it as proof of the existence of the regime in which general relativity breaks down. According to this viewpoint, the existence of singularities cannot be accepted 
\cite{wheeler, hawking3, bergmann}.
Another viewpoint, proposed by Misner
\cite{misner},
says that the general relativistic predictions of the singularity and its properties should be taken into account as it may tell us about what one should expect from some modification in the general theory of gravity which works in the regime of a strong field, for e.g., a quantum theory of gravity
\cite{burko}. 
Here, we consider the latter approach. In support of this approach, let us consider a scenario as follows: suppose we get some observational signatures from extreme gravity region. Now, these signatures contain traces of a quantum theory of gravity. If we already have the knowledge of the predictions of general relativity, then the difference in these observational traces and the predictions of general relativity may tell us how to tune general relativity so as to give the predictions which match with the observational signatures.

Once the existence of singularities is assumed, the next step is to comprehend the nature of the neighborhood of the singularity. One such property that needs to be investigated is the visibility of the singularity 
\cite{singh, psjoshi, harada, psjoshi2,  goswami0}. 
It is known that the big bang singularity is visible in principle because we can see the null and timelike geodesics coming from it.   However, it was still unclear whether or not the singularities arising as a result of gravitational collapse are necessarily censored completely from the outside universe by an event horizon. Penrose
\cite{penrose2}, 
in 1969, proposed what is now known as the cosmic censorship hypothesis, which is expressed in two forms: the weak cosmic censorship hypothesis \cite{tiplerclarkeellis}, which suggest that a singularity can never be globally naked, i.e., visible to faraway asymptotic observers.  Its strong counterpart, called the strong cosmic censorship hypothesis 
\cite{geroch, hawking2, penrose3},
states that the singularity can never be locally naked as well. Strong cosmic censorship is equivalent to a spacetime being globally hyperbolic 
\cite{joshinarlikar, psjoshi2}. 
It is a requirement for the uniqueness of the maximal global hyperbolic development of some initial data set. 
\cite{bruhat, ringstrom, sbierski}
One could refer to 
\cite{penrose4, israel, krolak2, christodoulou2, christodoulou3, christodoulou4, christodoulou5, christodoulou6} for further discussion on the development in the understanding of the cosmic censorship hypothesis.

It was, however, shown  by Eardley and Smarr
\cite{eardley}, 
Christodoulou
\cite{christodoulou},
and Joshi and Dwivedi
\cite{joshi}
that introducing inhomogeneity in the mass profile of the collapsing cloud could change the evolution of the apparent horizon, thereby possibly allowing nonspacelike geodesics to escape away from near the singularity without getting trapped. As such, the assumption of a homogeneous star is not very appropriate since it is expected that a star becomes denser as we move toward its center.  

Various mass distributions have been shown to give rise to visible singularities which are locally naked
\cite{goswami, joshi2, vaz, jhingan0}. 
The stability of locally naked singularities due to collapsing dust cloud against some perturbation in the initial data has also been studied by Deshingkar, Joshi, and Dwivedi
\cite{deshingkar}, 
and later by Mena, Tavakol, and Joshi
\cite{mena}.  
The local causal structure of the end state of the collapse in the presence of nonvanishing pressure has been studied wherein possibilities of locally naked singularities have been depicted \cite{magli1, magli2, giambo1, harada1, harada2, joshi7, mosani}. 

Nevertheless, the globally naked singularities, rather than the locally naked singularities, may have more observational significance. Some of the work dealing with global visibility can be found in \cite{deshingkar2, harada3, miyamoto, jhingan, kong, ortiz}. 
Deshingkar, Jhingan and Joshi 
\cite{deshingkar2} 
depicted some examples of mass functions giving rise to a globally visible singularity where the mass profile is a function of only $r$, i.e., the fluid under consideration was dust. The collapse, in this case, is considered to be marginally bound. Later, Jhingan and Kaushik
\cite{jhingan}
used a certain transformation of coordinates to put a restriction on the mass profile of a marginally bound collapsing dust to ensure global visibility of the singularity thus formed. 
On the contrary, Miyamoto, Jhingan and Harada \cite{miyamoto} investigated some stellar models (density distribution and total mass as the parameters) influenced by marginally stable configurations of neutron stars for various equations of state
\cite{shapiro}
and realized that for such configurations, the outgoing null geodesic, if at all it exists, gets trapped inside the event horizon, thereby making the singularity globally invisible.  Additionally, in massless scalar field collapse, even though naked singularities were shown to occur
\cite{Christodoulou7}, 
the initial data giving rise to such singularities have a positive codimension in a certain space of initial data. Hence, the singularity in such case is an unstable phenomenon, thereby preserving the cosmic censorship
\cite{Christodoulou8}.
Suggestions in support of the validity of weak cosmic censorship have also been discussed by Wald 
\cite{wald} and Hod \cite{hod}.

It is to be noted that the strength of singularities formed due to the depicted mass functions in \cite{deshingkar2, miyamoto, jhingan} 
was not investigated. If any object hits the singularity and is crushed to zero volume, then it is called a ``strong" singularity, according to Ellis and Schmidt
\cite{ellis}. 
The mathematically precise statement given by Tipler 
\cite{tipler} 
is as follows:

Let $\mathcal{M}$ be a smooth manifold of four dimensions along with a smooth metric $g$ with Lorentz signature $(-,+,+,+)$ defined on it. For a causal geodesic $\gamma: [t_0,0) \rightarrow \mathcal{M}$, the volume element defined by wedge product of three independent Jacobi field along $\gamma$, in a case $\gamma$ is a timelike geodesic (two independent Jacobi field in a case $\gamma$ is null geodesic), should approach to zero as $\lambda \rightarrow 0$, where $\lambda$ is the affine parameter along the geodesic.

We call such singularity as ``Tipler" strong. Sufficient condition for a singularity to be strong in this sense was provided by Clarke and Krolak 
\cite{clarke}.
Our basic purpose here is to examine the global causal structure of a singularity, keeping in mind the maintenance of its strength in the sense of Tipler, to ensure the physical relevance of the scenario considered. Also, marginally bound collapse is a very special case which corresponds to a very specific dynamics of the collapse, as we will see in the next sections. Considering such a collapsing scenario makes it easy to integrate one of Einstein's field equations. However, the generality is lost by doing so. Hence, we take into consideration here the nonmarginally bound collapse which incorporates all the possible dynamics of the collapse  (except one corresponding to marginally bound) depending on the functional form taken by the velocity function and permitted by the  Einstein's field equations, thereby widening our scope of understanding the gravitational collapse and its end state to a more general scenario.

The paper is arranged as follows: In Sec. II, Einstein's field equations corresponding to an inhomogeneous collapsing dust cloud is discussed. In Sec. III, the possibility of global visibility of singularities formed due to bound dust collapse is discussed.  In Sec. IV, the strength of such globally visible singularity in the sense of Tipler is discussed. We end the paper with the concluding remarks and stating a few open concerns in Sec. V. Here, we use the units in which $8\pi G=c=1$.

\section{Lemaitre-Tolman-Bondi spacetime}
The Lemaitre-Tolman-Bondi metric 
\cite{lemaitre, tolman, bondi} 
is a spherically symmetric metric governing the spacetime of collapsing dust clouds. It is given by
\begin{equation}
    ds^2=-dt^2+\frac{R'^2}{1+f}dr^2+R^2d\Omega^2
\end{equation}
in the comoving coordinates $t$ and $r$. We consider here a \textit{type I} matter field
\cite{hawking}.
In such a matter field, the energy-momentum tensor has nondiagonal entries as zero in a comoving coordinate system. One of the eigenvalues $\rho$ represents the energydensity as measured by a comoving observer at a point $p$. All observed fields with nonzero rest mass can be classified under type I matter field. The corresponding energy-momentum tensor along with vanishing pressure is given by
\begin{equation}
    T^{\mu \nu}=\rho U^{\mu} U^{\nu},
\end{equation}
where $U^{\mu}$, $U^{\nu}$ are the components of the four-velocity.
Einstein's field equations give us the expression of density and pressure, and the information about the dynamics of the collapse as
\begin{equation}\label{efe0}
    \rho=\frac{F'^2}{R^2R'},
\end{equation}
\begin{equation}\label{efe1}
    p=-\frac{\dot F}{R^2 \dot R},
\end{equation}
and
\begin{equation} \label{efe2}
    \dot R^2=\frac{F}{R}+f
\end{equation}
respectively. The superscripts dot and prime denote the partial derivative with respect to $t$ and $r$, respectively. Here, $F$ and $f$ are, respectively, called the Misner-Sharp mass function and the velocity function. The Misner-Sharp mass function in case of dust is a function of $r$ only and independent of $t$. This can be seen from Eq.(\ref{efe1}) which tells us that $\dot F=0$ since $p=0$ in case of dust. $F$ tells us about the mass of the collapsing cloud inside a shell of radial coordinate $r$ at time $t$. For zero pressure, this mass is conserved inside a fixed radial shell. For the collapsing matter field to be well behaved at the initial time and at the center of the cloud, certain regularity conditions need to be maintained. The metric functions should by $\mathcal{C}^2$ differentiable everywhere according to the obligations of the Einstein's field equations. The Misner-Sharp mass function should have the following expression:
\begin{equation}\label{regularitycondition}
    F(r)=r^3 M(r).
\end{equation}
Here, $M>0$ and is a regular, at least $\mathcal{C}^2$ function, having a finite value at the limit of approach to the center, and is called the mass profile of the collapsing cloud. In the case $F$ goes as $r^2$ or lower power, it could be seen from the Einstein's field equation (\ref{efe0}) that the density blows up at the center at the initial epoch itself, which is undesirable. Additionally, in order to avoid cusp in the energy density, the function space of $M$ is further restricted to follow the condition 
\begin{equation}
    M'(0)=0.
\end{equation}
The positivity of energy density is achieved by restricting $F'>0$ and $R'>0$. This is maintained by restricting the mass profile as follows:
\begin{equation}
    3M+rM'>0.
\end{equation}
Energy density can also be positive when $F'<0$ and $R'<0$. However, in such a case, the mass profile becomes negative near the center, which is not allowed.

A collapsing solution of Einstein's field equation is obtained by restricting the physical radius as $\dot R<0$. This means that a particular shell of fixed radial coordinate collapses to form a singularity when $R=0$ for this shell. However, $R$ vanishes also at the regular center. Both these cases can be differentiated by expressing the physical radius as
\begin{equation}
    R(t,r)=r v(t,r).
\end{equation}
Now, a shell of radial coordinate $r$ is said to form a singularity a time $t_s$ when $v(t_s,r)=0$. Rescaling of the physical radius is done using the coordinate freedom such that
\begin{equation}\label{scaling freedom}
    R(t_i,r)=r,
\end{equation} 
where $t_i$ is the initial time. This can be rewritten as $v(t_i,r)=1$.

The polarity of $f$ classifies the spacetime in three different categories: bound (elliptic), marginally bound (flat) and unbound (hyperbolic) collapse, corresponding to the restrictions $f<0$, $f=0$ and $f>0$ respectively. Rewriting Eq.(\ref{efe2}) as $\dot v=\sqrt{M+f/r^2}$ demands that the velocity function should have the form
\begin{equation}
    f=r^2b_0(r)
\end{equation}
as a regularity requirement. Here, $b_{0}(r)$ is a sufficiently differentiable function.

Equation (\ref{efe2}) can be integrated to get 
\begin{equation}\label{timecurve}
    t-t_s(r)=-\frac{R^{\frac{3}{2}}G(-fR/F)}{\sqrt{F}}.
\end{equation}
Here $G(y)$ is defined as follows:
\begin{equation}\label{G}
    \begin{split}
        & G(y)= \left(\frac{\text{arcsin}\sqrt{y}}{y^{\frac{3}{2}}}-\frac{\sqrt{1-y}}{y} \right) \hspace{0.5cm} \text{for} \hspace{0.5cm} 0<y<1, \\
        & G(y)=\frac{2}{3} \hspace{0.5cm} \text{for} \hspace{0.5cm} y=0, \\
        & G(y)= \left(\frac{-\text{arcsinh}\sqrt{-y}}{(-y)^{\frac{3}{2}}}-\frac{\sqrt{1-y}}{y} \right) \hspace{0.5cm} \text{for}  -\infty<y<0. 
    \end{split}
\end{equation}
The constant of integration in Eq.(\ref{timecurve}) can be obtained using Eq.(\ref{scaling freedom}) as
\begin{equation}\label{singularitycurve}
     t_s(r)=\frac{r^{3/2}G(-fr/F)}{\sqrt{F}}.
\end{equation}
This is called the singularity curve. It gives us the information about the time at which a shell of radial coordinate $r$ collapses to form a singularity $R=0$. 

A singularity can be only locally naked if the null geodesic can escape from the neighborhood of the singularity but later in its path, comes across the trapped surfaces, and falls back to the singularity. The boundary of all trapped surfaces is called the apparent horizon. The evolution of the apparent horizon is determined by equating the physical radius with the Misner-Sharp mass function as $R=F(r)$. This, along with Eqs.(\ref{timecurve}) (\ref{singularitycurve}) gives us the time of formation of the apparent horizon as a function of radial coordinate as
\begin{equation}\label{ahcurve}
    t_{AH}(r)=\frac{r^{3/2}G(-rf/F)}{\sqrt{F}}-FG(-f).
\end{equation}
It is also called the apparent horizon curve. 
$F(0)=0$ implies $t_s(0)=t_{AH}(0)$, thereby creating a possibility for nonspacelike geodesic to have a positive tangent at $r=0$. Such singularities are at least locally naked. The geodesics may later get trapped, thereby keeping the weak cosmic censorship intact. However, it is also possible that the singular geodesic avoids getting trapped by the trapped surfaces and reaches the boundary of the collapsing cloud unhindered. Such singularities are studied in detail in the next section.

\section{Global visibility}

\begin{figure*}\label{fig1}
\subfigure[]
{\includegraphics[scale=0.5]{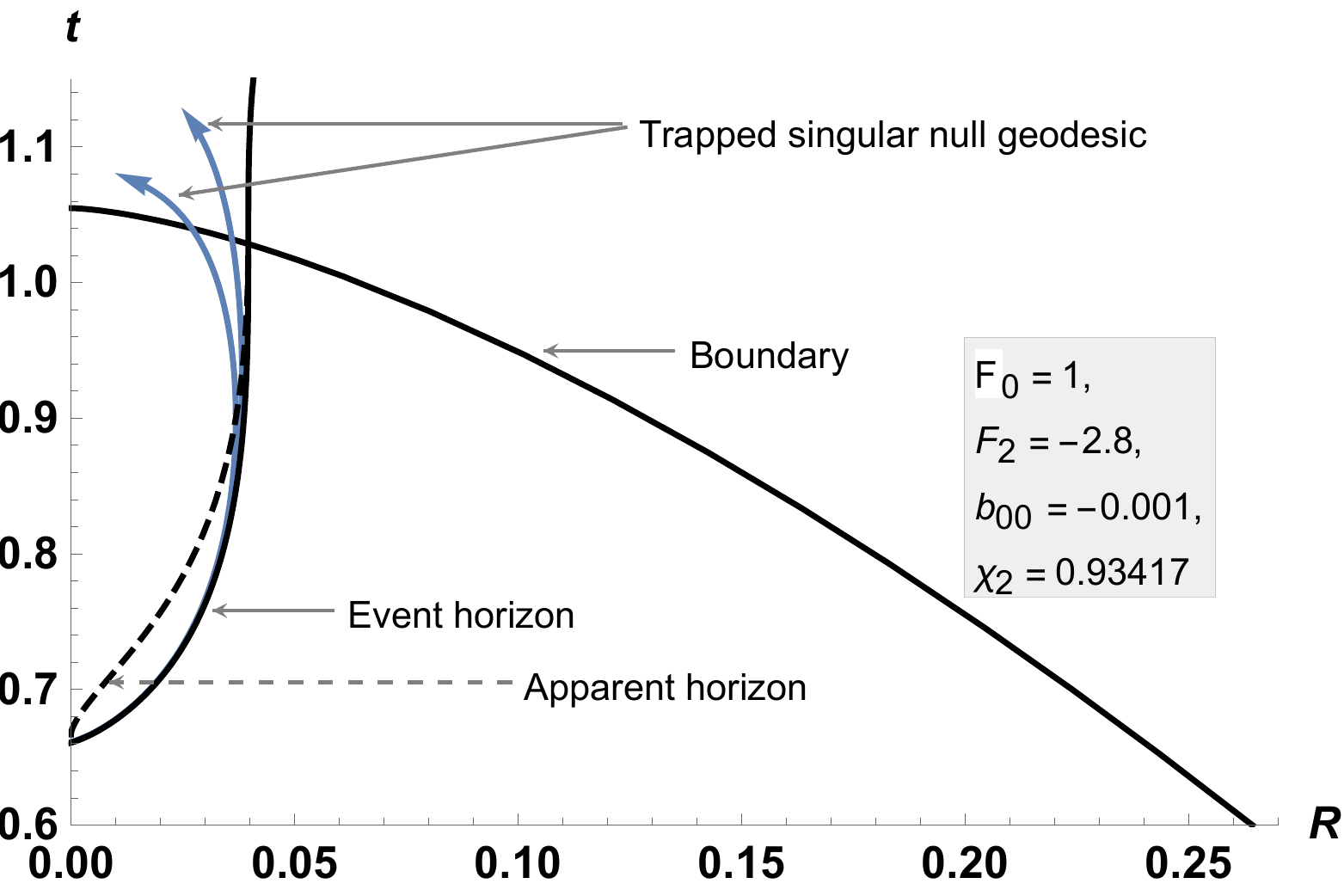}}
\hspace{0.2cm}
\subfigure[]
{\includegraphics[scale=0.5]{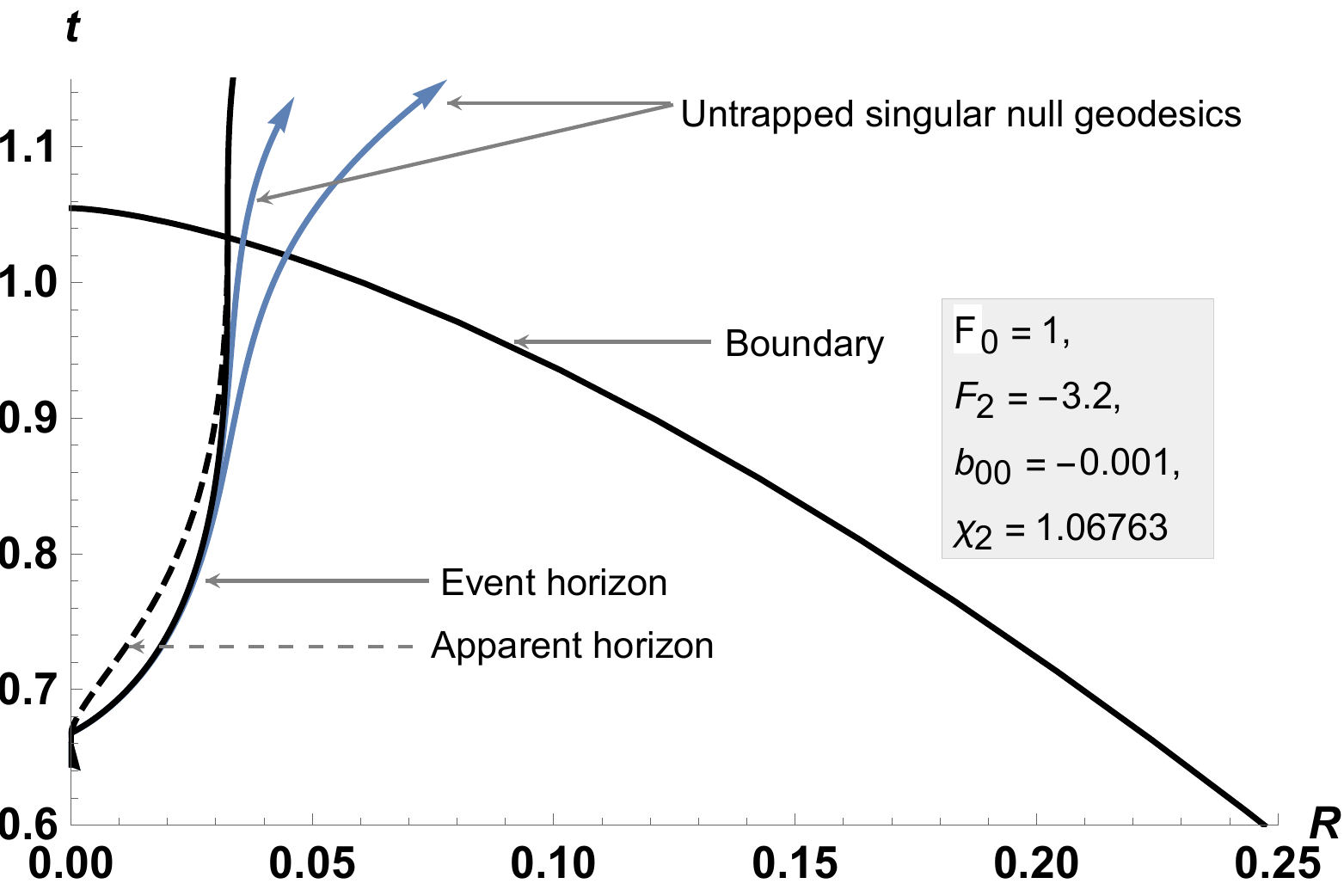}}
\caption{Causal structure of a singularity formed as an end state of a bound (elliptic) collapsing dust cloud. Apparent horizon, event horizon, and singular null geodesics are represented by dashed black curves, solid black curves, and
solid blue curves, respectively. (a) The evolution of the event horizon starts from the center before the formation of the central singularity. Singular null geodesics, if at all, can escape the singularity gets trapped later and falls back in, making the singularity only locally naked. (b) The evolution of the event horizon starts during the formation of the central singularity. Singular null geodesics can escape and reach the faraway observer. Here, $\frac{fR}{F}\sim10^{-3}$ initially, and reduces in magnitude thereafter, in both these cases. Higher-order terms: $o(y_1^3)$ and $o(y_1^2)$, arising in Eq.(\ref{R(t,r)}) are neglected.} 
\end{figure*}
The singularities which are only locally visible may not be of much observational significance. This is because, in such a case, an observer outside the event horizon will not be able to receive any signal escaping from the neighborhood of the singularity. For this reason, it is of extreme importance to investigate whether or not there exists a globally visible singularity. 

For a singularity to be globally visible, null geodesics originating from the neighborhood of the singularity should not only avoid getting trapped by the trapped surfaces but also reach the boundary of the star before the event horizon. It turns out that for globally visible singularity, the latter always implies the former. This is because, at $r=r_c$, the apparent horizon coincides with the event horizon. We know that the evolution of the event horizon of the collapsing cloud is the same as the evolution of the null geodesic along with the condition that at the boundary of the cloud $r_c$, the following equality should be satisfied:
\begin{equation}\label{icforeh}
    F(r_c)=R(t,r_c).
\end{equation}
Now, the event horizon cannot start forming after the initiation of the formation of trapped surfaces (or its boundary, i.e., the apparent horizon). This is because any null geodesic, more specifically outgoing null geodesic, forming inside the apparent horizon, will have a negative tangent and fall back into the singularity. The evolution of EH can be thought of as the evolution of the last outgoing radial null geodesic escaping the center without getting trapped and falling back to the singularity. The equation of the null geodesic is given by
\begin{equation}\label{NG}
    \frac{dt}{dr}=\frac{R'}{\sqrt{1+f}}.
\end{equation}
In the case of inhomogeneous dust, at $r=0$, the time of formation of AH is the same as the time of formation of central singularity, as seen from Eq.(\ref{singularitycurve}) and Eq.(\ref{ahcurve}) and the fact that $F$ vanishes. Hence, it can be concluded that the  EH starts forming either before or during the formation of the  singularity due to the collapse of the central shell i.e.
\begin{equation}
t_{EH}(0)\leq t_s(0).    
\end{equation} 
Here, $t_{EH}(r)$ is the event horizon curve which is the solution of Eq.(\ref{NG}) with the condition given by Eq.(\ref{icforeh}).
 
We now define a small neighborhood around the time of formation of central singularity such that the null geodesic escaping the center at a time belonging to this neighborhood (an interval around $t_s(0)$ rather than a single point $t_s(0)$) will be termed singular. This neighborhood should have a size of the order of Planck time.

One may question the choice of the size of this neighborhood as a magnitude influenced by quantum theory, even when general relativity is assumed to be fundamental. Let us recall that, as mentioned in the Introduction; we interpret the singularity theorem as proposed by Misner
\cite{misner}.
Hence, we will determine the result obtained by the general relativistic approach in the strong gravity regime, which may help us to predict what we must expect from a quantum theory of gravity.

If we can trace a singular null geodesic (SNG) reaching the boundary before the event horizon, then we have 
\begin{equation}
 t_{SNG}(r_c)<t_{EH}(r_c).   
\end{equation}
Now, we know that $R$ is a monotone decreasing function of $t$ since $\dot R<0$. Hence we have
\begin{equation}
R( t_{SNG}(r_c),r_c)>R(t_{EH}(r_c),r_c).
\end{equation}
However, we know that $R(t_{EH}(r_c),r_c)=F(r_c)$ from Eq.(\ref{icforeh}). Hence for a singular null geodesic reaching the boundary before the event horizon, the following inequality should be satisfied:
\begin{equation}
    R(t,r_c)>F(r_c).
\end{equation}
Geometrically, the above inequality gives a positive value of the expansion parameter for outgoing null geodesic congruence $\Theta_l$, at $r=r_c$, which is expressed in terms of physical radius, Misner-Sharp mass function and velocity function as follows:
\begin{equation}
   \Theta_l= \frac{2}{R}\left( \sqrt{1+f}-\sqrt{\frac{F}{R}+f} \right).
\end{equation}
This specifies the divergent nature of these outgoing null geodesic congruences at $r=r_c$.

Now, the expression of $R$ in terms of the comoving coordinates $t$ and $r$ is obtained from Eqs.(\ref{timecurve}) and (\ref{singularitycurve}) as 
\begin{equation}\label{Rfornonzerof}
    R=\left(\frac{r^{\frac{3}{2}}G\left(-f r/F\right)-\sqrt{F}t }{G\left(-f R/F\right)}\right)^{\frac{2}{3}}.
\end{equation}
In the case of marginally bound collapse, this is reduced to 
\begin{equation}\label{R(t,r)mb}
    R=\left(r^{\frac{3}{2}}-\frac{3}{2}\sqrt{F}t \right)^{\frac{2}{3}}.
\end{equation}
However, in the case of nonmarginally bound collapse, we use the Taylor expanded expression for the function $G(y)$ given by Eq.(\ref{G})  around $y=0$ for $0<y\leq 1$ as
\begin{equation}\label{taylorexpandedG}
    G(y)=\frac{2}{3}+\frac{1}{5}y+\frac{3}{28}y^2+o(y^3).
    \end{equation}
This  can then be used in Eq.(\ref{Rfornonzerof}) to write $R$ explicitly as
\begin{widetext}
\begin{equation}\label{R(t,r)}
    R(t,r)=\frac{5F}{2f}\left(1-\sqrt{1- o(y_1^3)-\frac{4f}{5F}\left(r^{\frac{3}{2}}\left(1-\frac{3fr}{10F}+o(y_2^2)\right)-\frac{3}{2}\sqrt{F}t\right)^{\frac{2}{3}}}\right),
\end{equation}
\end{widetext}
for nonvanishing velocity function, i.e. $f\neq 0$. 
Here, 
\begin{equation}
    y_1=-\frac{fR}{F}, \hspace{1cm} y_2=-\frac{fr}{F}.
\end{equation}
Ignoring higher order, i.e. $o(y_1^3)$ and $o(y_2^2)$ in Eq.(\ref{R(t,r)}) is equivalent to considering the expansion of $G$ from Eq.(\ref{taylorexpandedG}) only up to first order. Hence, large value of the ratio $\frac{fR}{F}$ may not give a good approximation. Therefore, in our investigation, we make sure to keep this ratio small by considering positive velocity function having small deviation from zero. 

Deshingkar, Jhingan, and Joshi \cite{deshingkar2} studied the global causal structure of the end state of marginally bound collapse, wherein three different mass distributions were considered. These mass distributions had first, second, and third-order inhomogeneity terms, respectively, in the initial density. (Here, $nth$ order inhomogeneity term means the initial density profile is of the form $\rho(r)=\rho_0+\rho_n r^n$, where $\rho_0$ and $\rho_n$ are constants. Also, the corresponding Misner-Sharp mass function is of the form $F=F_0r^3+F_{n+3} r^{n+3}$ where $F_0$ and $F_{n+3}$ are constants). The general result obtained was that a higher magnitude of the inhomogeneity term corresponded to the end state as a globally visible singularity. Here, we analyze the global behavior of the singularity formed by bound collapse and for a mass function and the velocity function given by
\begin{equation}
 F=F_0r^3+F_2r^5,\hspace{1cm} f=b_{00}r^2
\end{equation}
The boundary of the cloud is found such that the density smoothly matches to zero there. Hence, the boundary is given by
\begin{equation}
    r_c=\sqrt{-\frac{3F_0}{5F_2}}.
\end{equation}
 This is a second-order inhomogeneity in the mass function.  As seen in Figure 1 the singularity is at least locally naked for chosen values of $F_0$ and $F_2$. However, in Figure 1(a) the event horizon starts forming before the formation of the central singularity, thereby making the singularity globally hidden. The singular geodesic can escape the singularity but later gets trapped and falls back. Now, increasing the magnitude of the inhomogeneity term $F_2$, as seen in Figure 1(b) affects the evolution of the event horizon in such a way that its time of formation is delayed and now overlaps with the time of formation of a central singularity. A null geodesic with the property $F(r_c)<R(t,r_c)$ can be traced with the criteria that the difference between the time of escape of the null geodesic from the center and the time of formation of the central singularity can be reduced as much as we desire. In such a case, the singularity is considered as globally visible.

It should be noted that even if such globally visible singularities exist, it should not create a problem for the cosmic censorship if such singularities are gravitationally weak. We discuss this in more detail in the following section.

\section{Strength of the singularity}


\begin{figure*}\label{fig2}
\subfigure[]
{\includegraphics[scale=0.5]{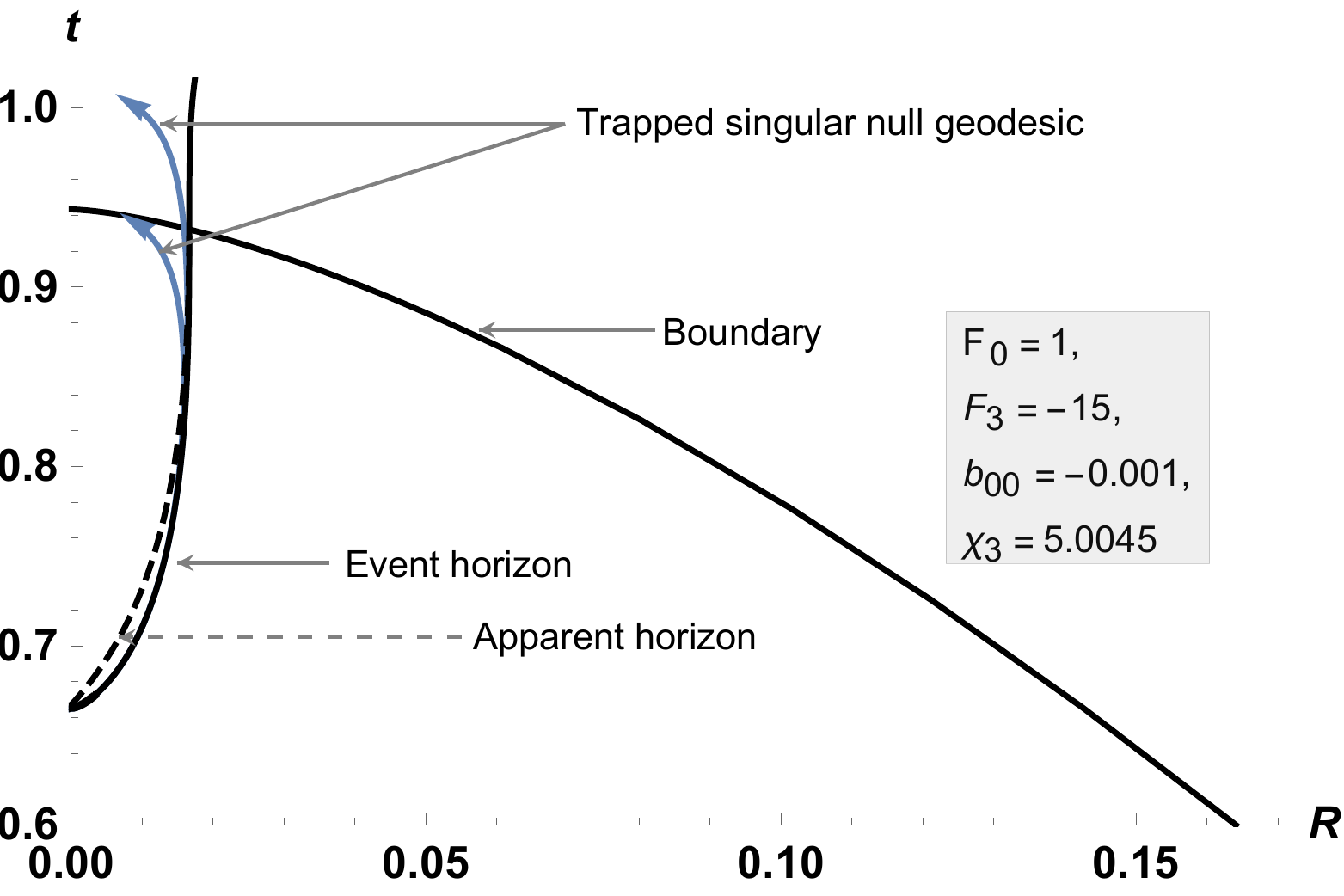}}
\hspace{0.2cm}
\subfigure[]
{\includegraphics[scale=0.5]{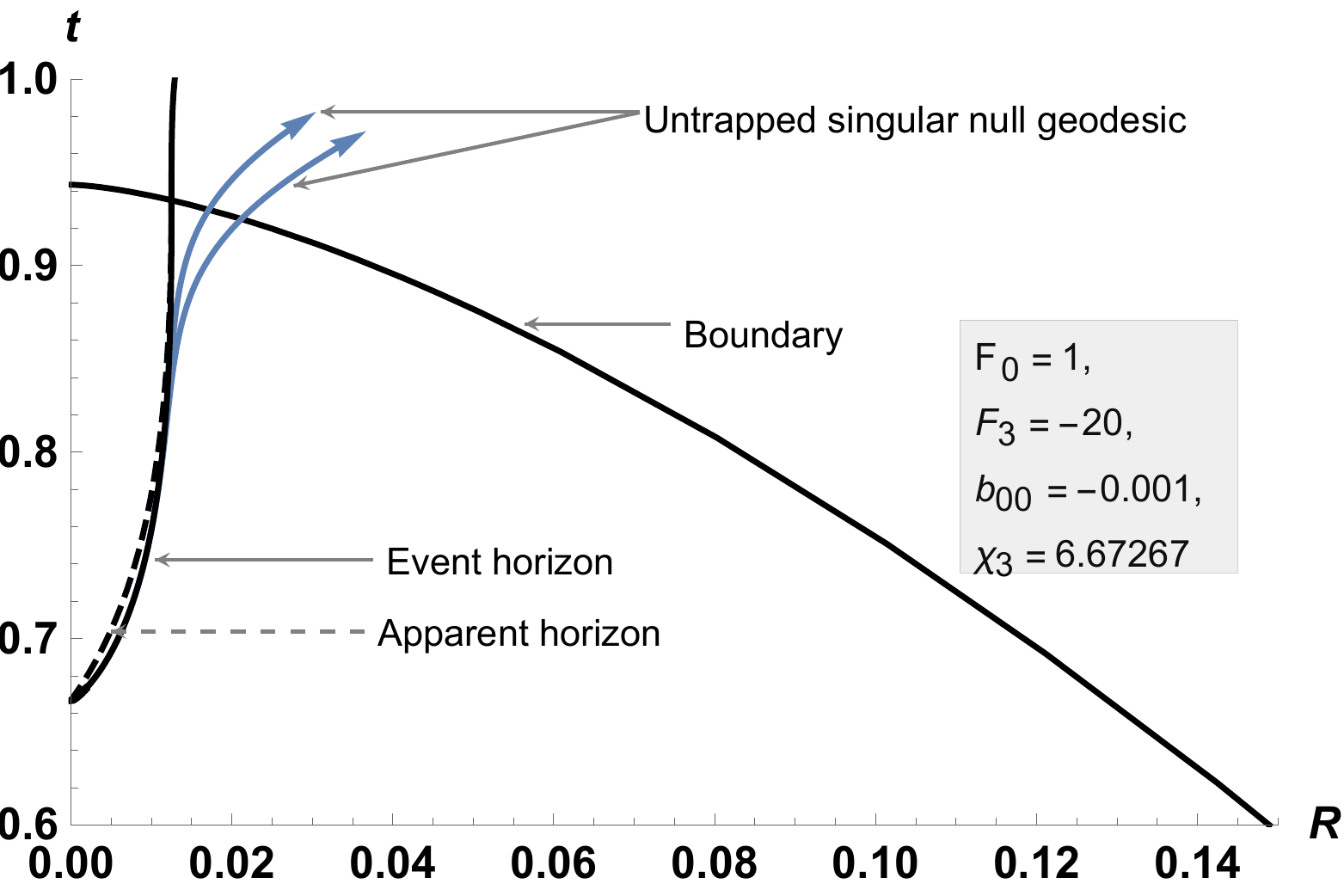}}
\caption{Causal structure of a Tipler strong singularity formed as an end state of a bound (elliptic) collapsing dust cloud. Apparent horizon, event horizon, and singular null geodesics are represented by dashed black curves, solid black curves, and
solid blue curves, respectively. $\chi_1$ = $\chi_2$ = 0 and  $\chi_3>0$. (a) The evolution of the event horizon starts from the center before the formation of the central singularity. Singular null geodesics, if at all, can escape the singularity gets trapped later and falls back in, making the singularity only locally naked. (b) The evolution of the event horizon starts during the formation of the central singularity. Singular null geodesics can escape and reach the faraway observer. Here, $\frac{fR}{F}\sim10^{-3}$ initially, and reduces in magnitude thereafter, in both these cases. Higher-order  terms: $o(y_1^3)$ and $o(y_1^2)$, arising in Eq.(\ref{R(t,r)}) are neglected.}
\end{figure*}

\begin{figure}\label{unbound}
\includegraphics[scale=0.5]{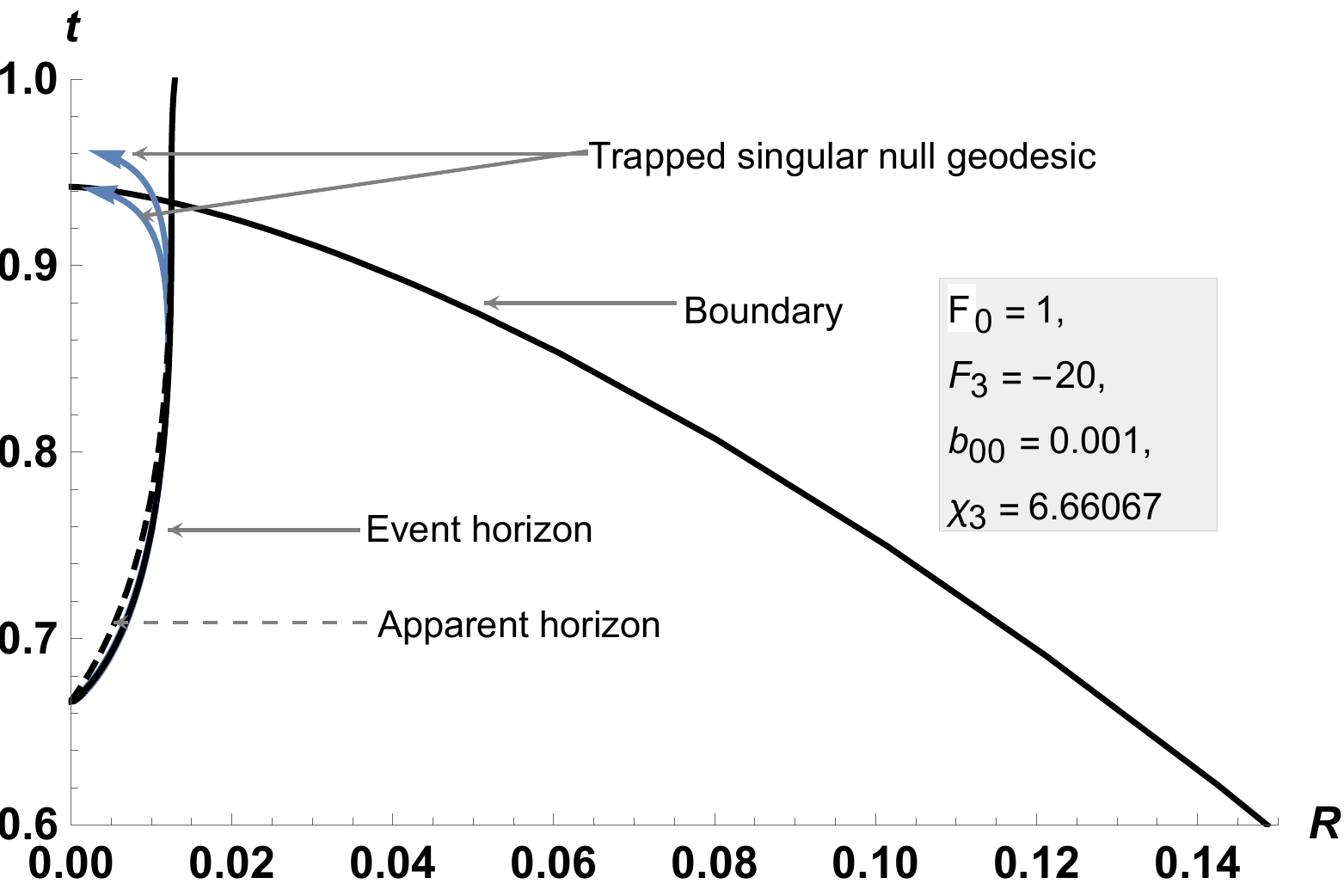}
\caption{Causal structure of a Tipler strong singularity formed as an end state of an unbound (hyperbolic) collapsing dust cloud. Apparent horizon, event horizon, and singular null geodesics are represented by dashed black curves, solid black curves, and solid blue curves, respectively.    $\chi_1=\chi_2=0$ and $\chi_3>0$. The mass profile, which ends as a globally visible singularity in bound case (see Figure (2)), ends as a globally hidden singularity in unbound case. Here, $\frac{fR}{F}\sim10^{-3}$ initially, and reduces in magnitude thereafter. Higher-order  terms: $o(y_1^3)$ and $o(y_1^2)$, arising in Eq.(\ref{R(t,r)}) are neglected.}
\end{figure}

To maintain the strength of the singularity in the sense of Tipler, Clarke, and Krolak has given a necessary and sufficient condition which needs to be satisfied. The criterion says that at least along one null geodesic, the following inequality needs to be satisfied:

   \begin{equation} \label{Krolak and Clarke criteria}
    \lim_{ \lambda \to 0} \lambda ^2R_{ij}K^iK^j>0.
\end{equation}
Here, $\lambda$ is the affine parameter along the null geodesic with $\lambda=0$ at the singularity. We can use this criterion to put a restriction on a particular parameter signifying the nonlinear relation between the physical radius and the tangent of the outgoing radial null geodesic at the singular center.  The time curve can be Taylor expanded around the center $r=0$ as follows:
\begin{equation}\label{taylorsinglaritycurve}
    t(r,v)=t(0,v)+r\chi_1(v)+r^2\chi_2(v)+r^3\chi_3(v)+O(r^4),
\end{equation}
where
\begin{equation}\label{chii}
    \chi_i(v)=\frac{1}{i!}\frac{d^i t}{dr^i}\bigg |_{r=0}.
\end{equation}    

For a singularity to be at least locally visible, the tangent of the future directed radial null geodesic from the singularity at $r\to 0$ should be positive. In the $(R,u)$ frame, where $u=r^{\alpha}$ with $\alpha>1$, this tangent is written as $X_0=\lim_{r\to 0}\frac{dR}{du}$.    
It can be shown that 
\begin{equation}\label{X0}
\begin{split}
X_{0}^{\frac{3}{2}}= & \lim_{r\to 0}\frac{1}{\alpha-1}\big ( \chi_1(0) +2r\chi_2(0)+3r^2\chi_3(0) \\
& +4r^3\chi_4(0)+o(r^4)\big ) \sqrt{M_0(0)}r^{\frac{5-3\alpha}{2}}.
\end{split}
\end{equation}
Here, the relation between the tangent of outgoing radial null geodesic at the singularity and the components $\chi_i$ of the Taylor expansion of the time curve at $v=0$ is depicted. To ensure the positivity of $X_0$, the first nonzero $\chi_i$ should be positive.

Now, it is known that Eq.(\ref{Krolak and Clarke criteria}) can be satisfied only if  $\alpha \geq 3$. Also, the necessary criterion for the singularity to be at least locally naked is given by $\alpha \leq 3$. Hence, the necessary criterion for a singularity to be strong and locally naked is given by 
\cite{joshi} 
\begin{equation}\label{alpha=3}
    \alpha=3.
\end{equation}

\begin{figure*}\label{fig3}
{\includegraphics[scale=0.315]{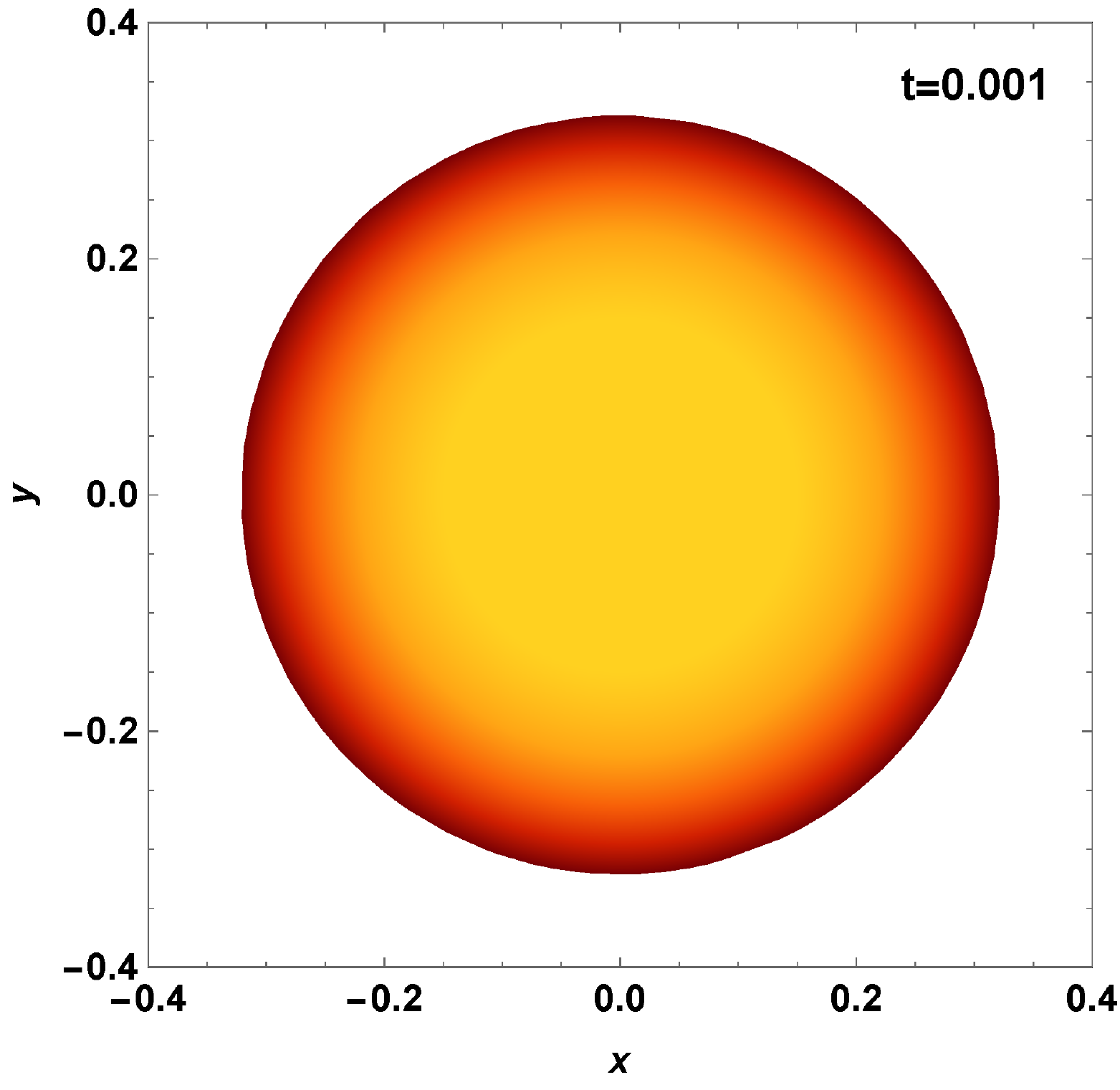}\includegraphics[scale=0.315]{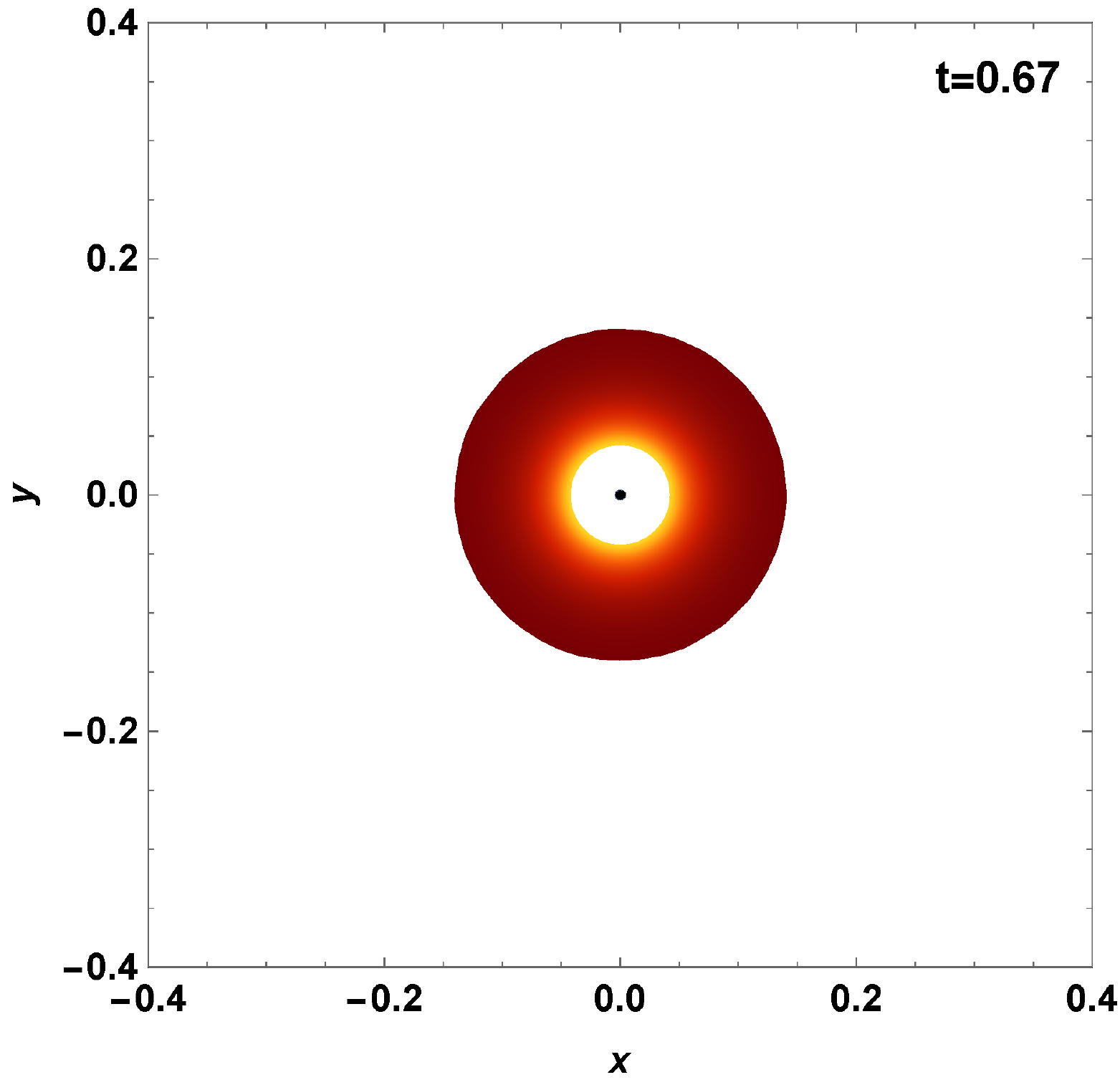}\includegraphics[scale=0.315]{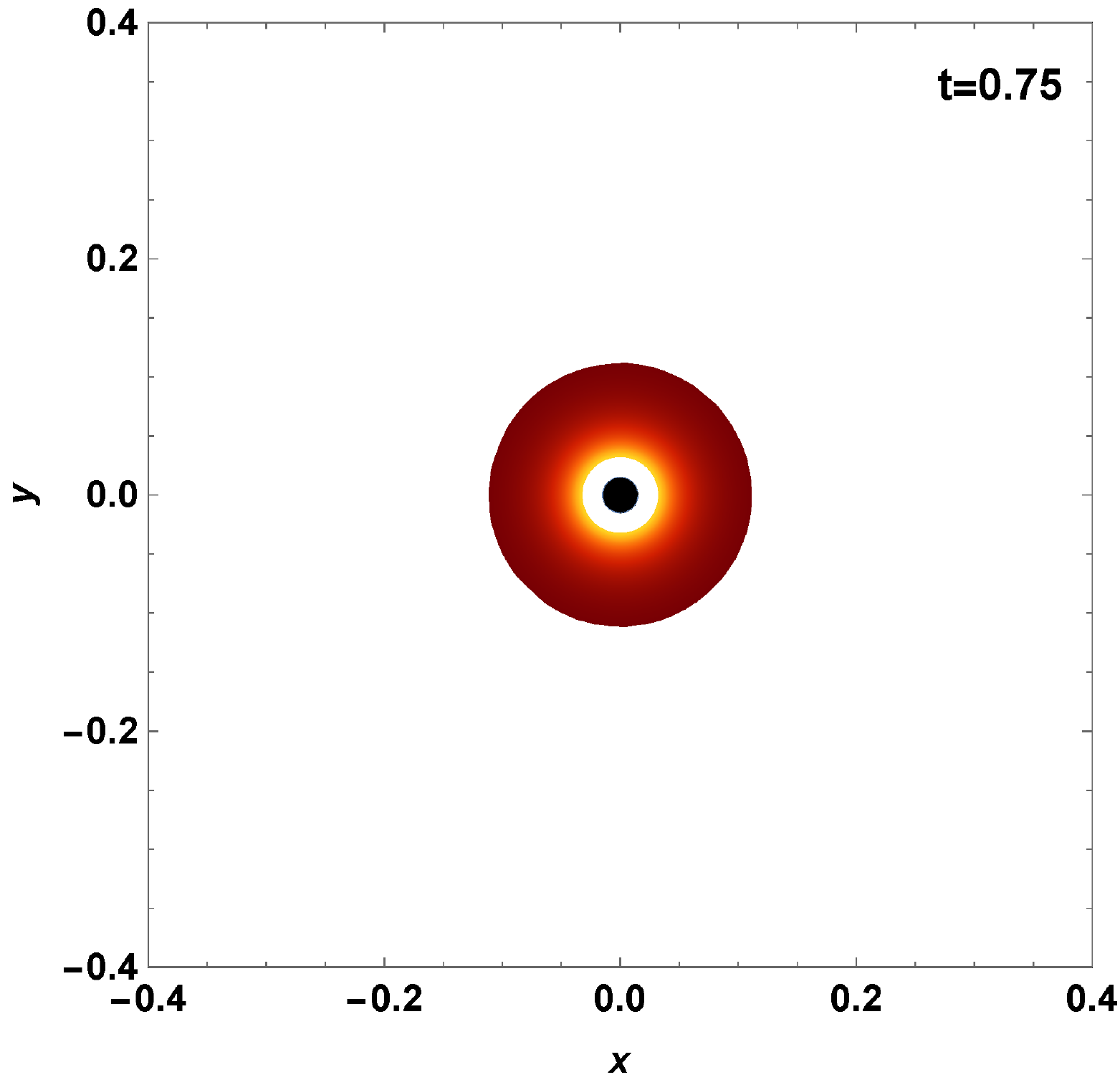}}
{\includegraphics[scale=0.315]{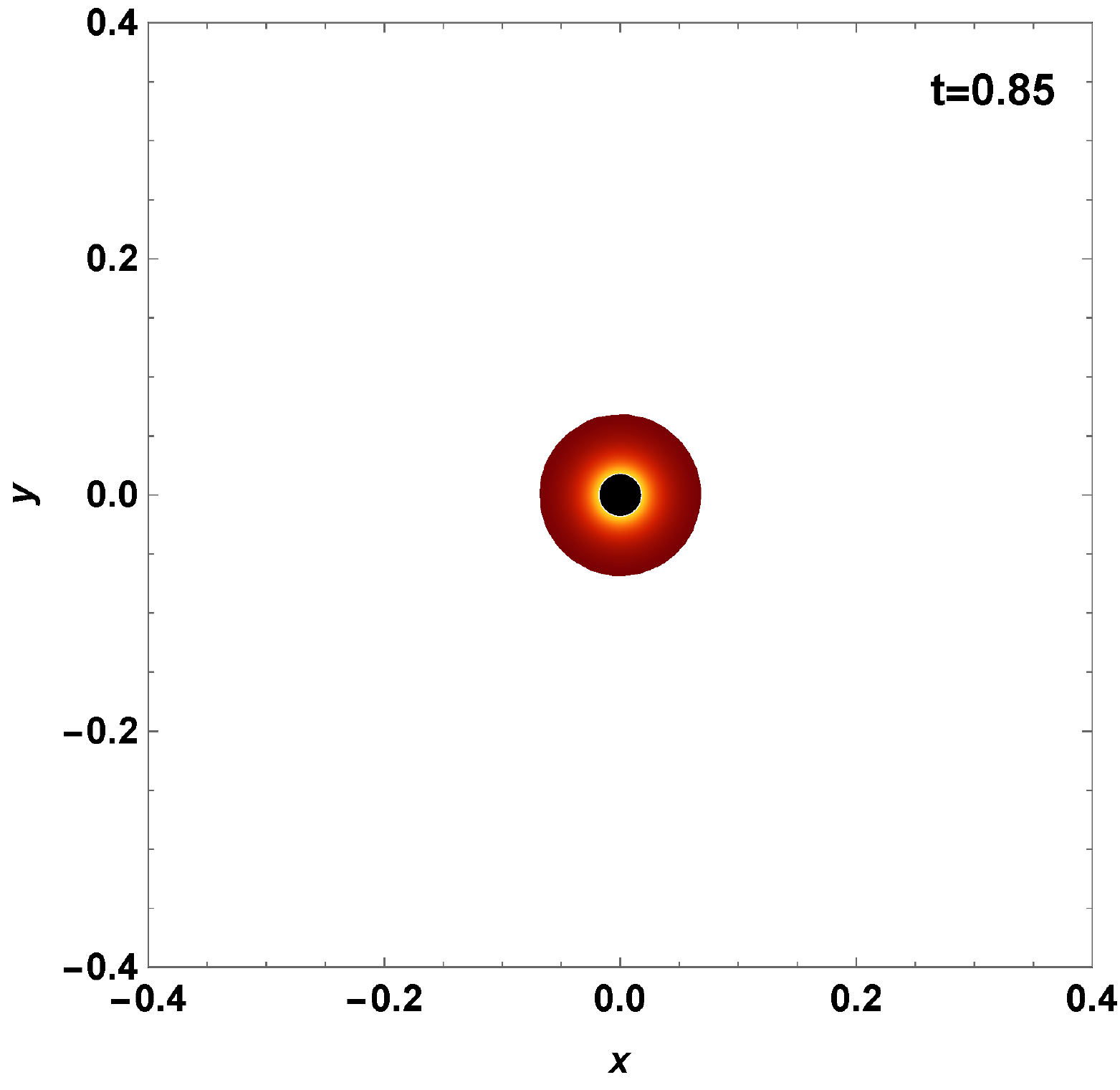}\includegraphics[scale=0.315]{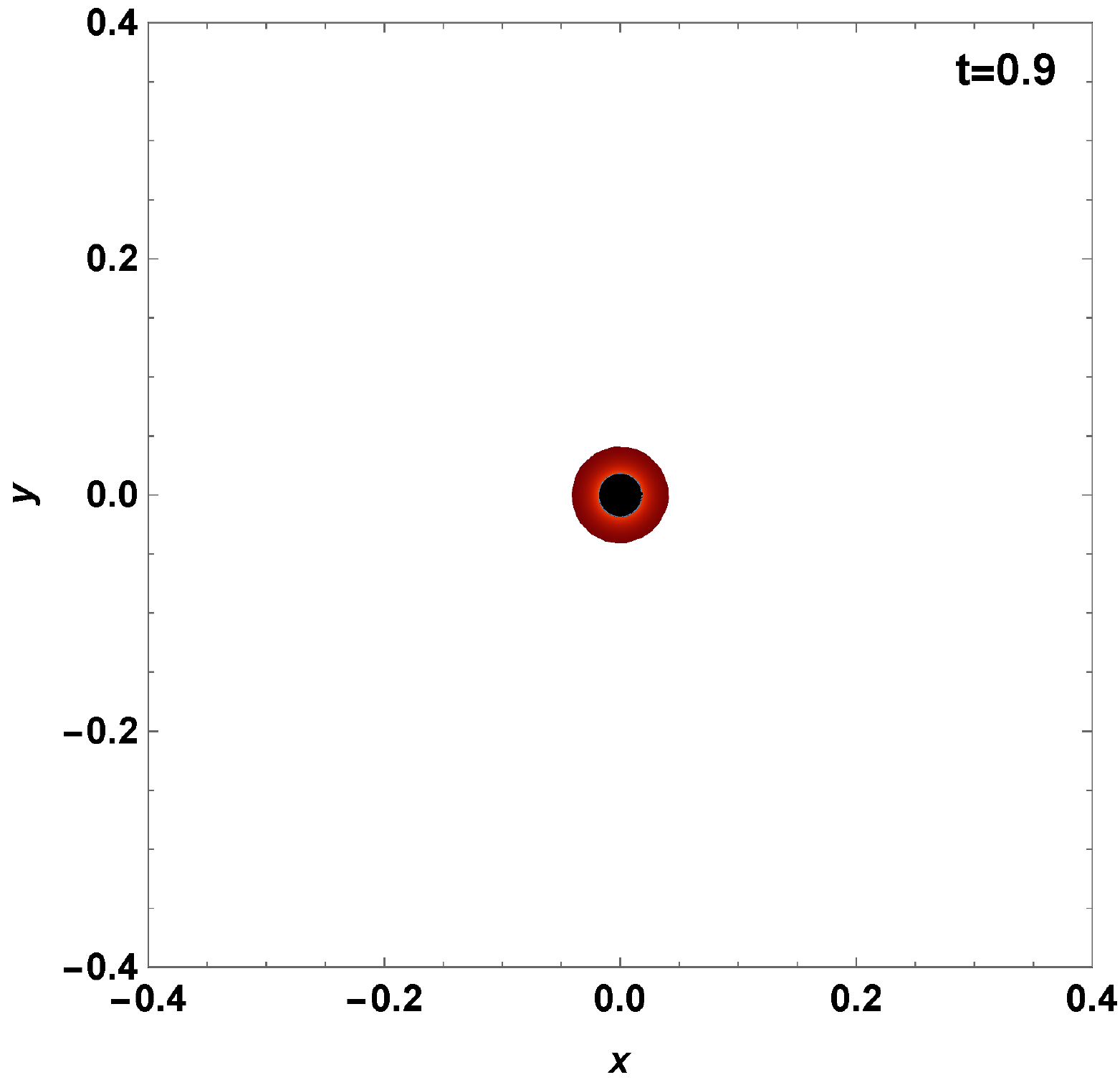}\includegraphics[scale=0.315]{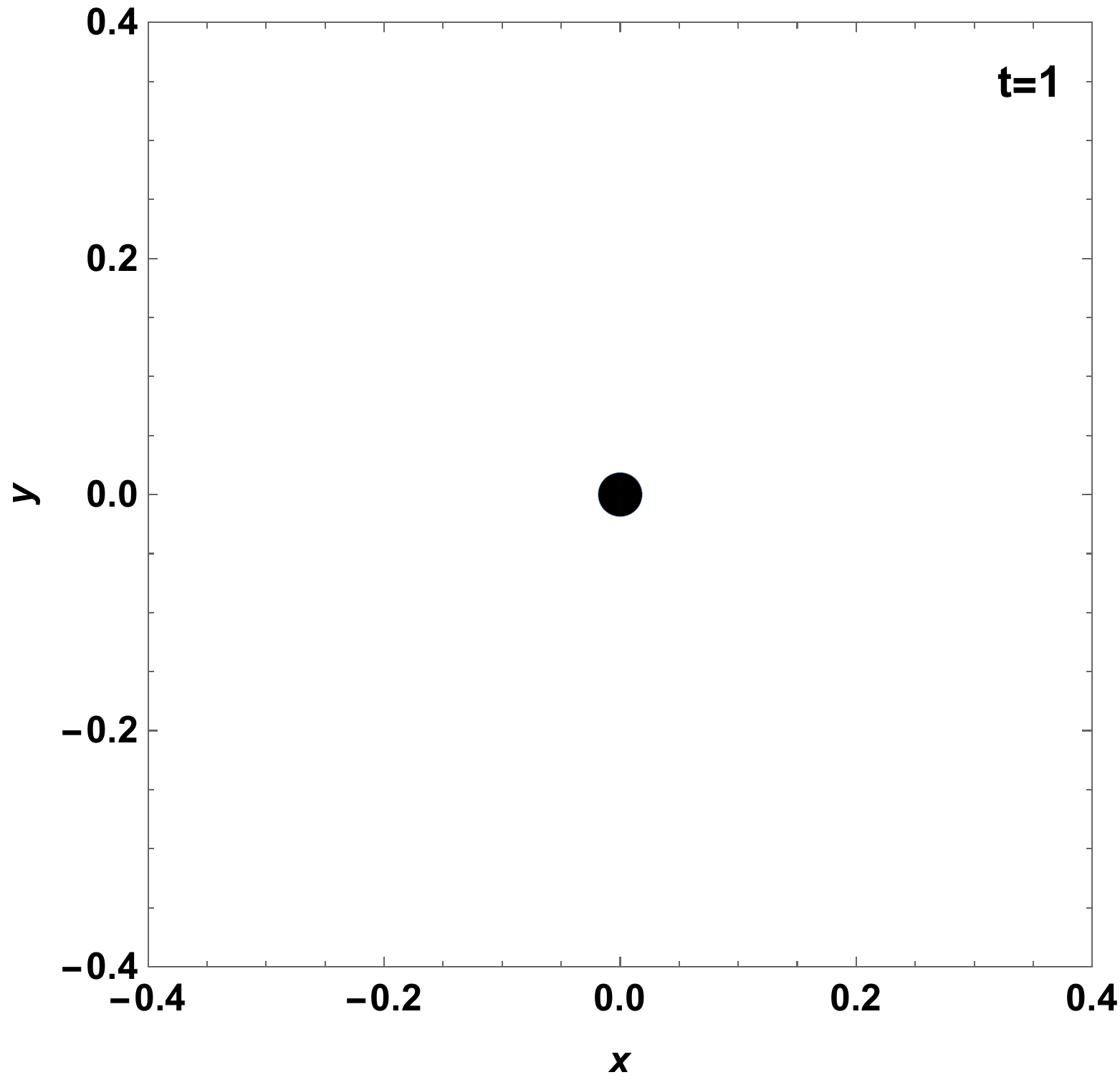}}
\caption{Evolution of the collapsing star and the global causal structure is depicted here. $F_0=1$, $F_3=-15$ and $F_i=0$ for $i \neq 1,3$. $b_{00}=-0.001$ and $b_{0j}=0$ for $j \neq 0$. $\frac{fR}{F}\sim10^{-3}$ initially and reduces in magnitude thereafter. Higher-order  terms: $o(y_1^3)$ and $o(y_1^2)$, arising in Eq.(\ref{R(t,r)}) are neglected.} The singularity is Tipler strong with $\chi_1=\chi_2=0$ and $\chi_3\neq 0$.  The solid black disk represents the event horizon which increases in size with time. No singular geodesic can escape and reach the boundary.
\end{figure*}
\begin{figure*}\label{fig4}
{\includegraphics[scale=0.315]{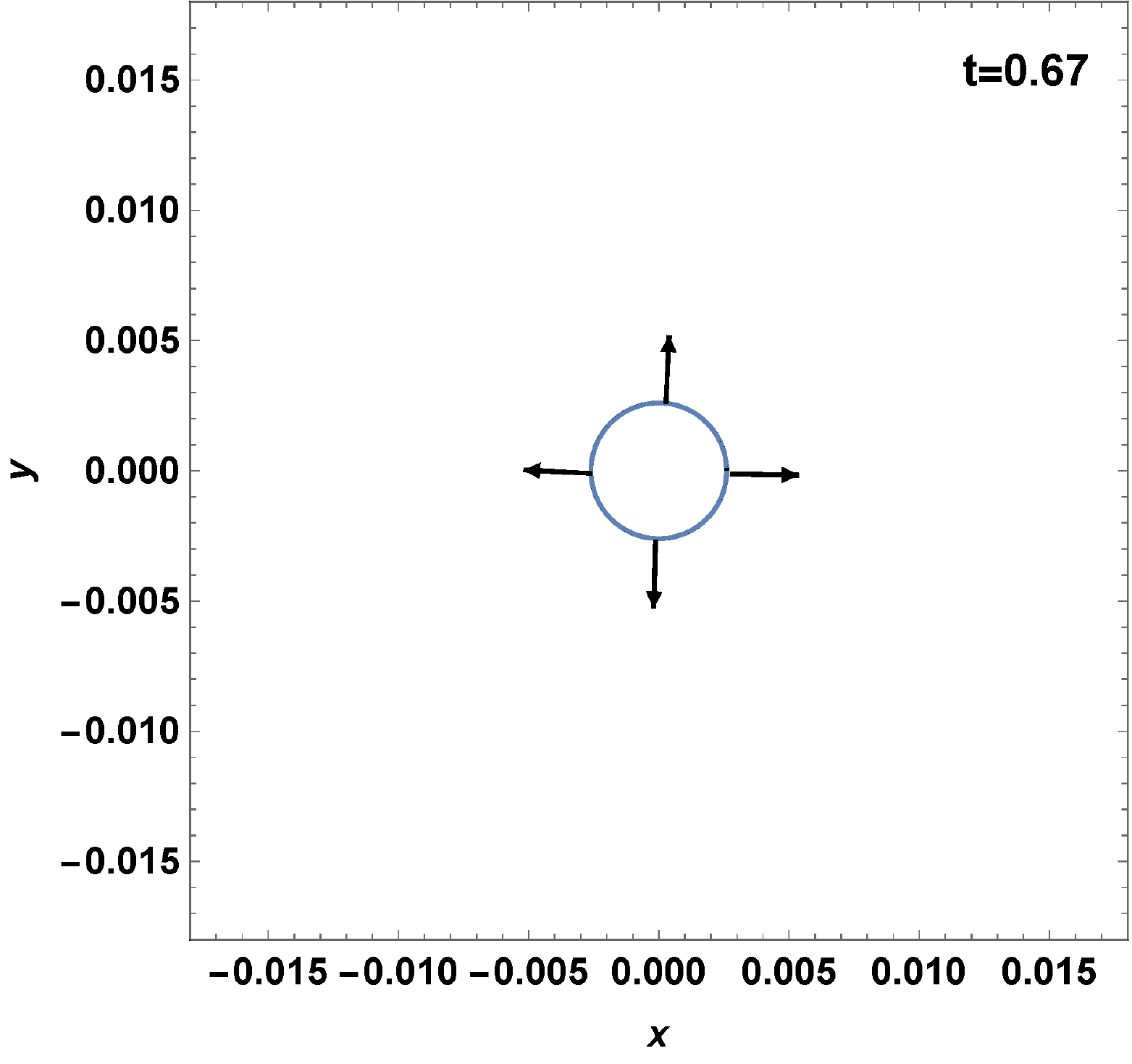}\includegraphics[scale=0.315]{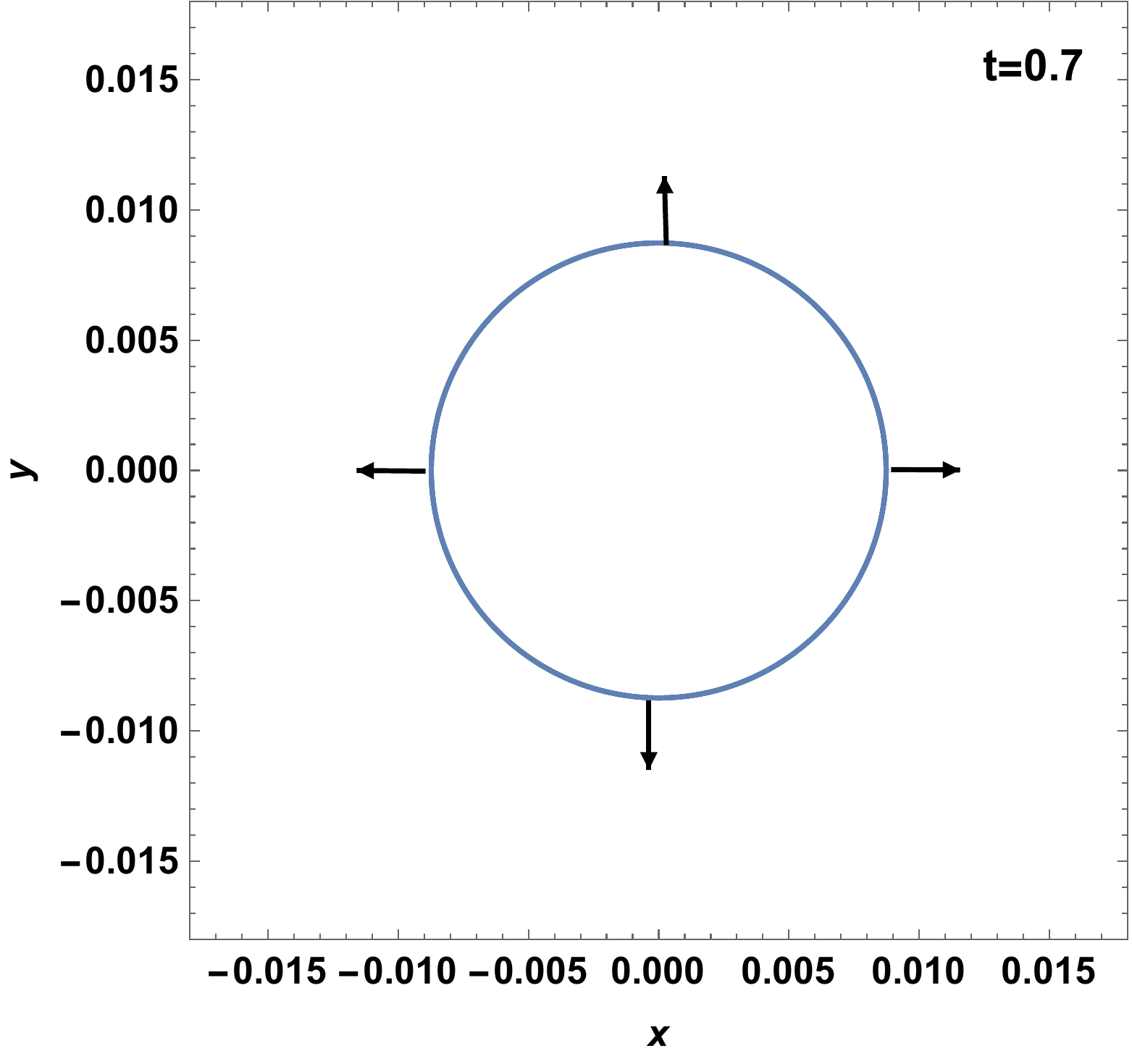}\includegraphics[scale=0.315]{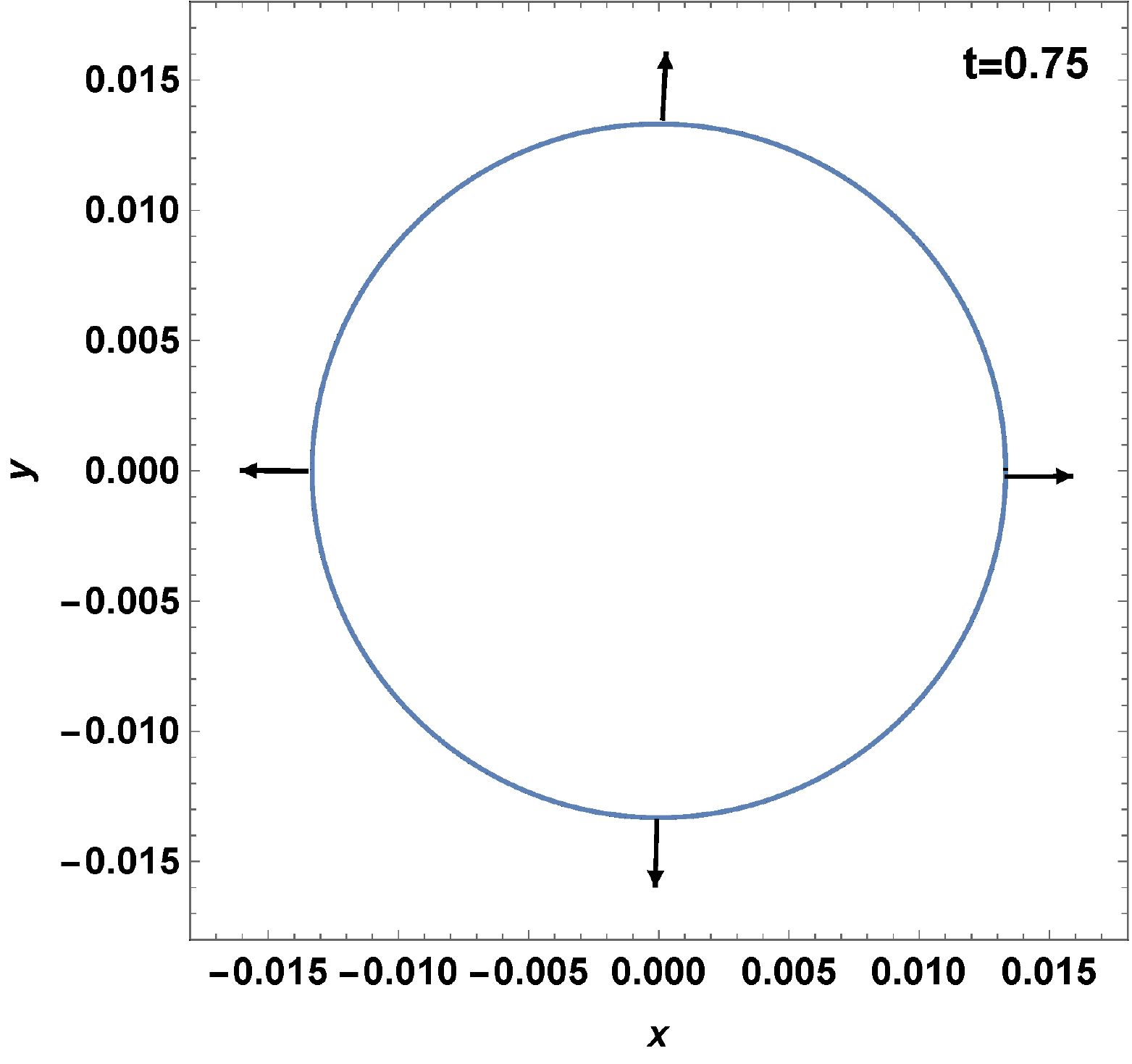}}
{\includegraphics[scale=0.315]{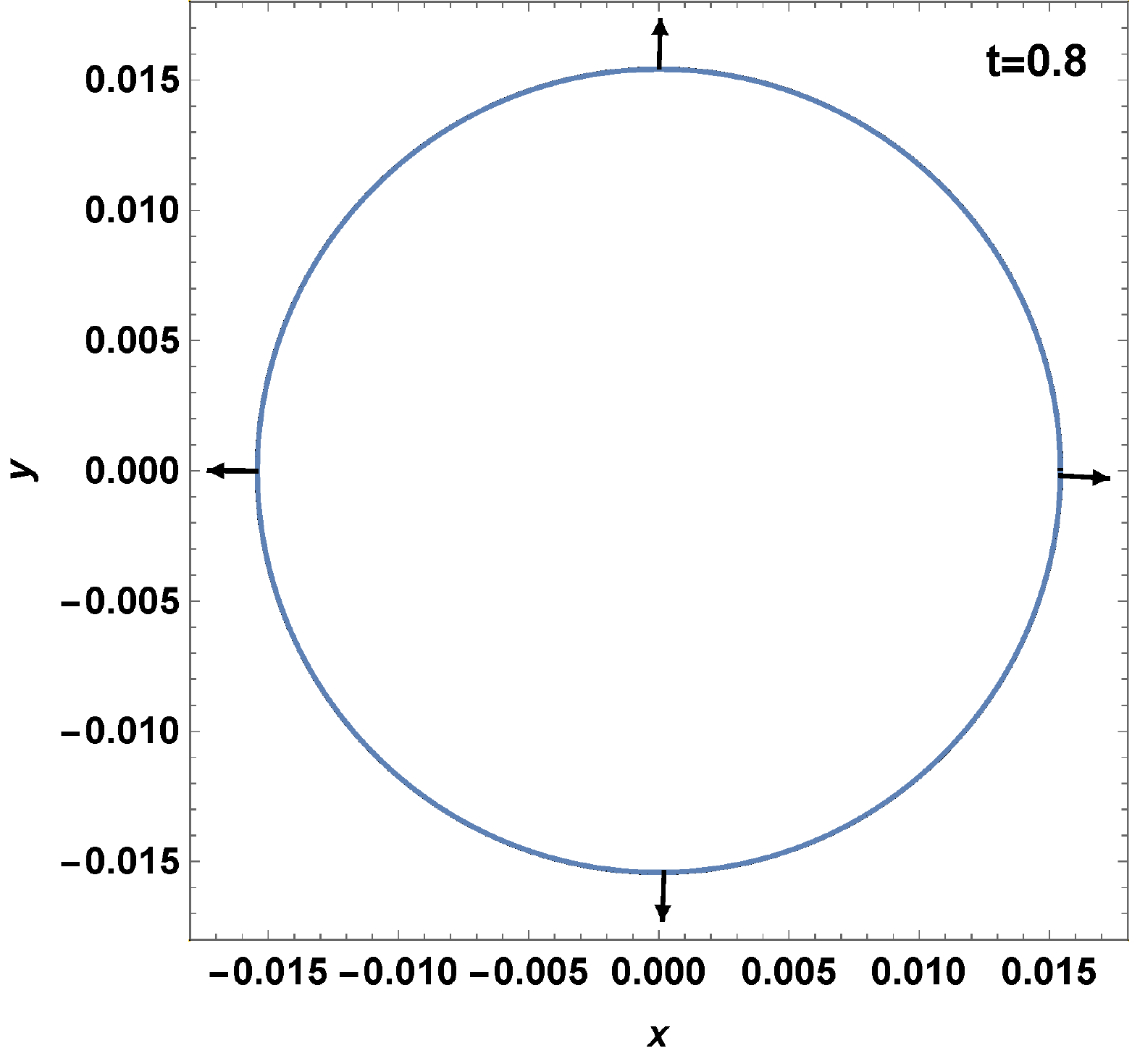}\includegraphics[scale=0.315]{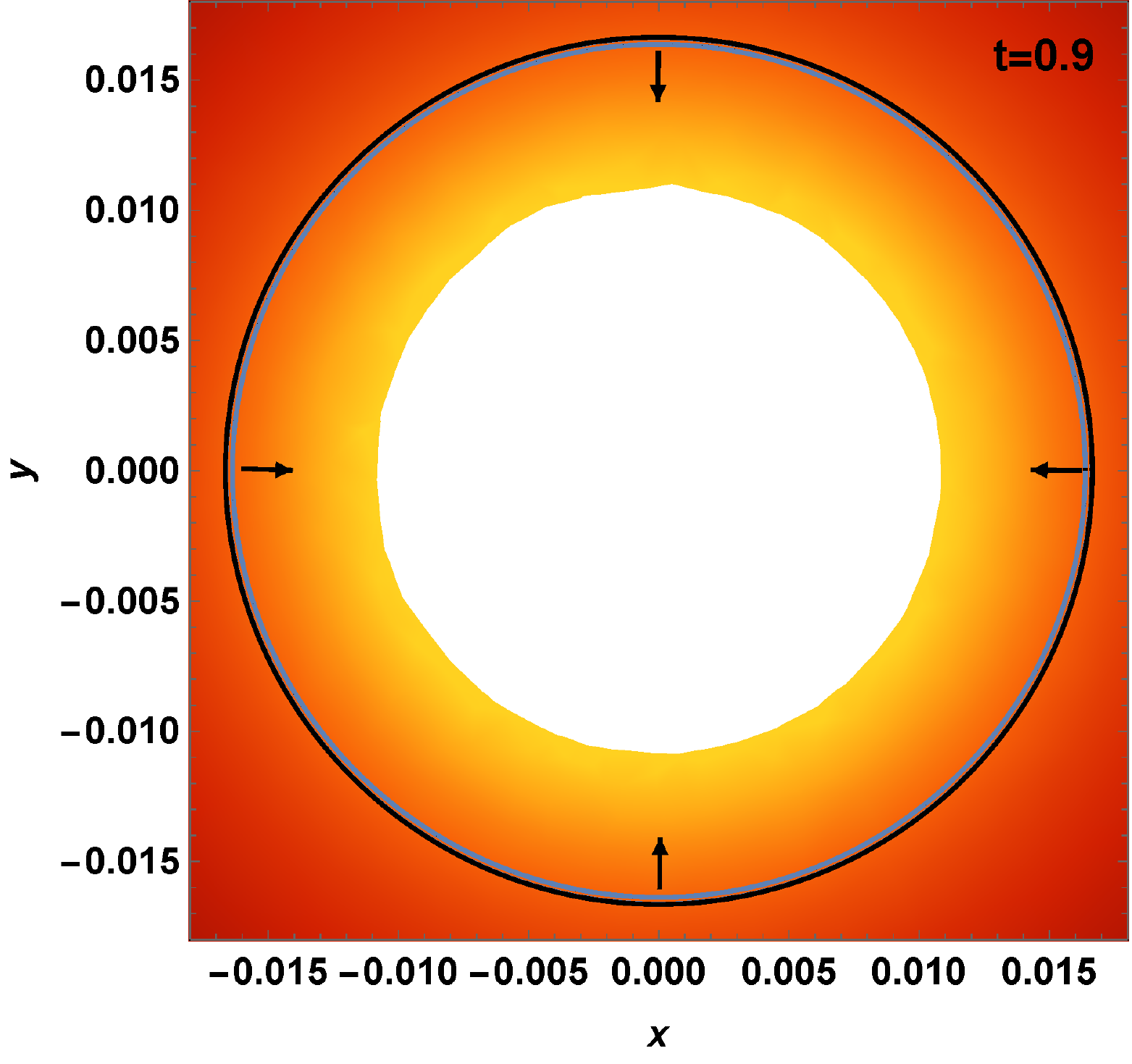}\includegraphics[scale=0.315]{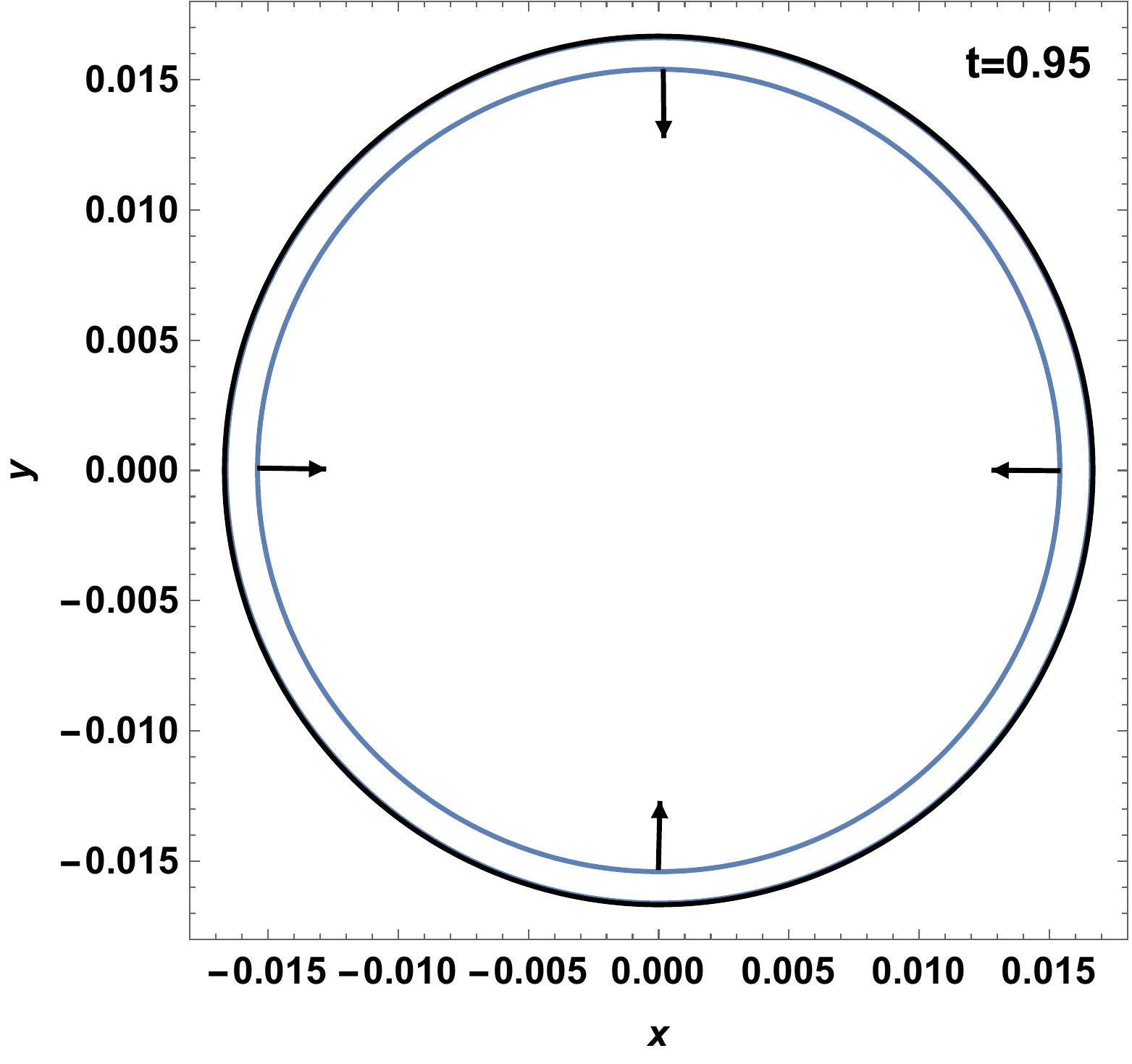}}
\caption{Local causal structure is depicted here. $F_0=1$, $F_3=-15$ and $F_i=0$ for $i \neq 1,3$. $b_{00}=-0.001$ and $b_{0j}=0$ for $j \neq 0$. $\frac{fR}{F}\sim10^{-3}$ initially and reduces in magnitude thereafter. Higher-order  terms: $o(y_1^3)$ and $o(y_1^2)$, arising in Eq.(\ref{R(t,r)}) are neglected.} The singularity is Tipler strong with $\chi_1=\chi_2=0$ and $\chi_3\neq 0$. Behavior of singular outgoing radial null geodesic wave front is represented by blue color. Event horizon is represented by black colored circle.
\end{figure*}
\begin{figure*}\label{fig5}
{\includegraphics[scale=0.315]{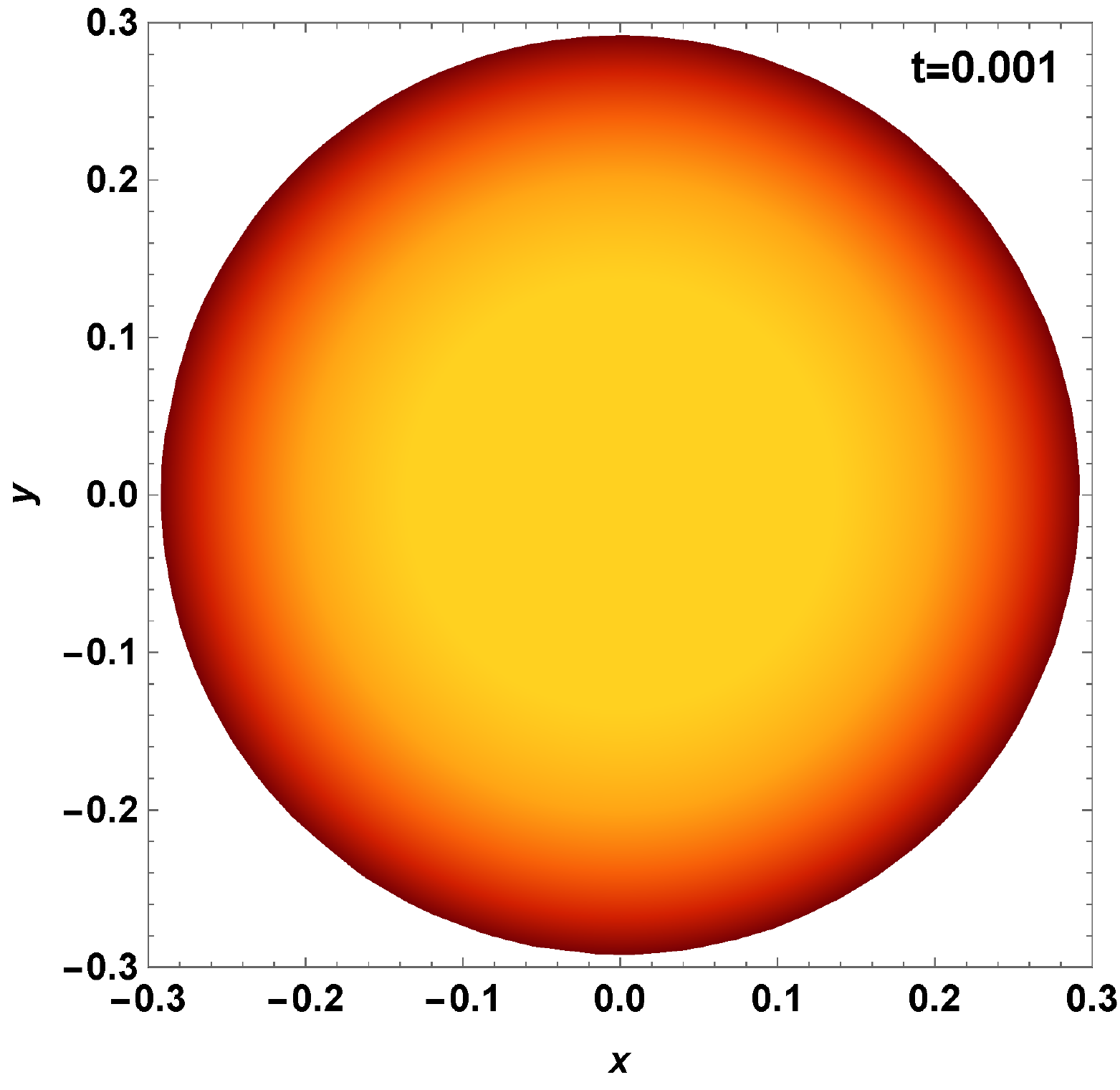}\includegraphics[scale=0.315]{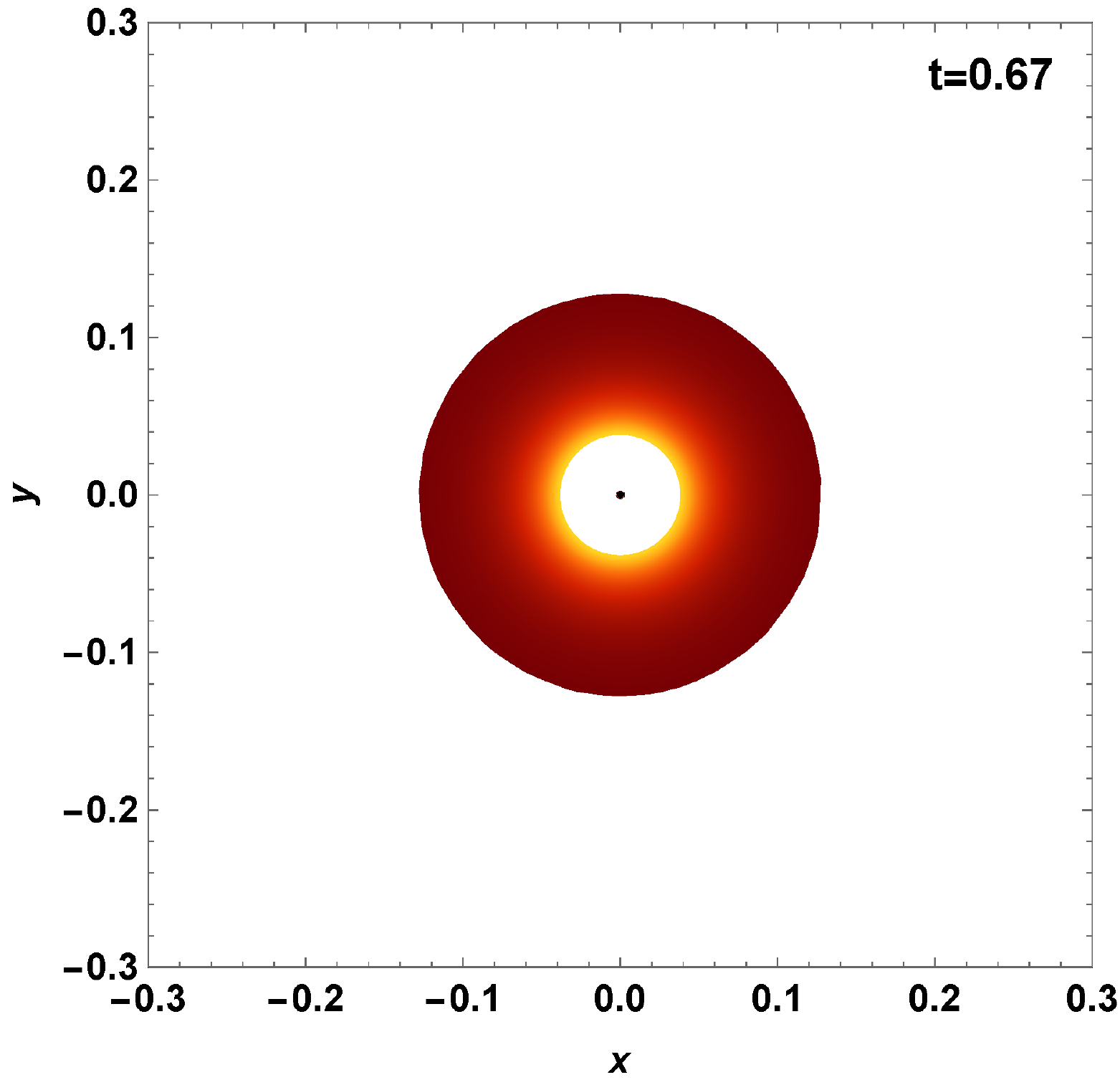}\includegraphics[scale=0.315]{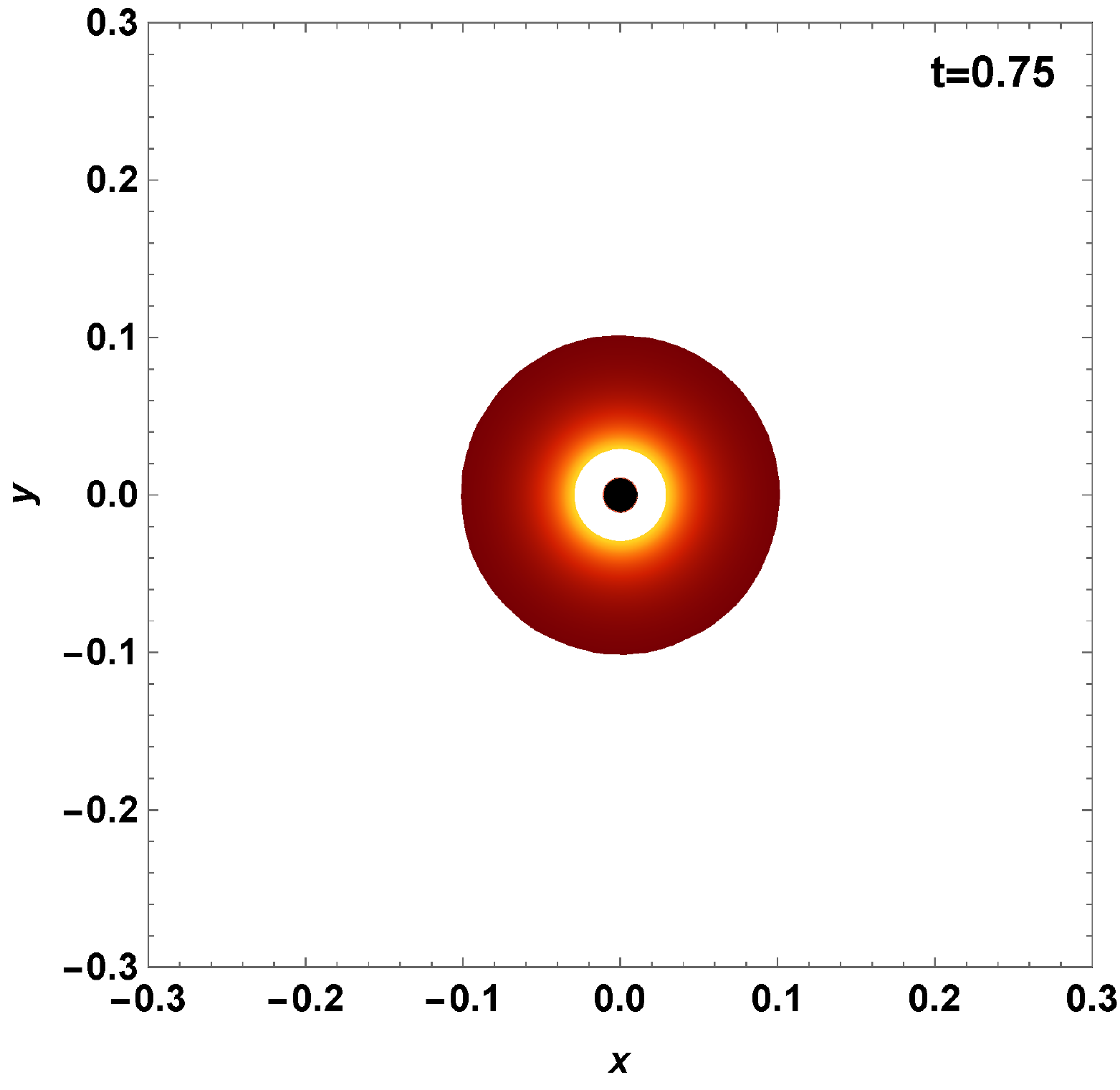}}
{\includegraphics[scale=0.315]{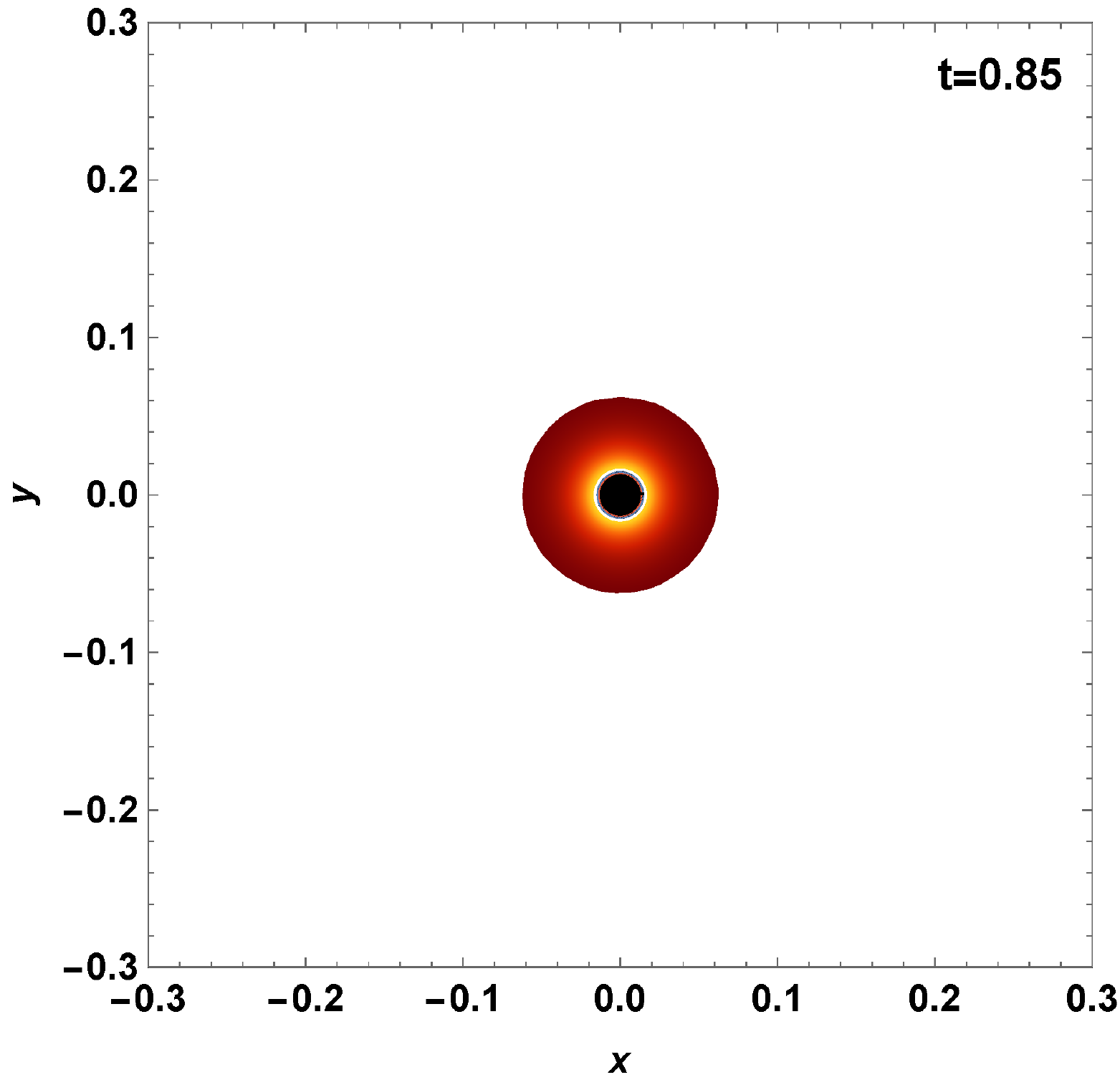}\includegraphics[scale=0.315]{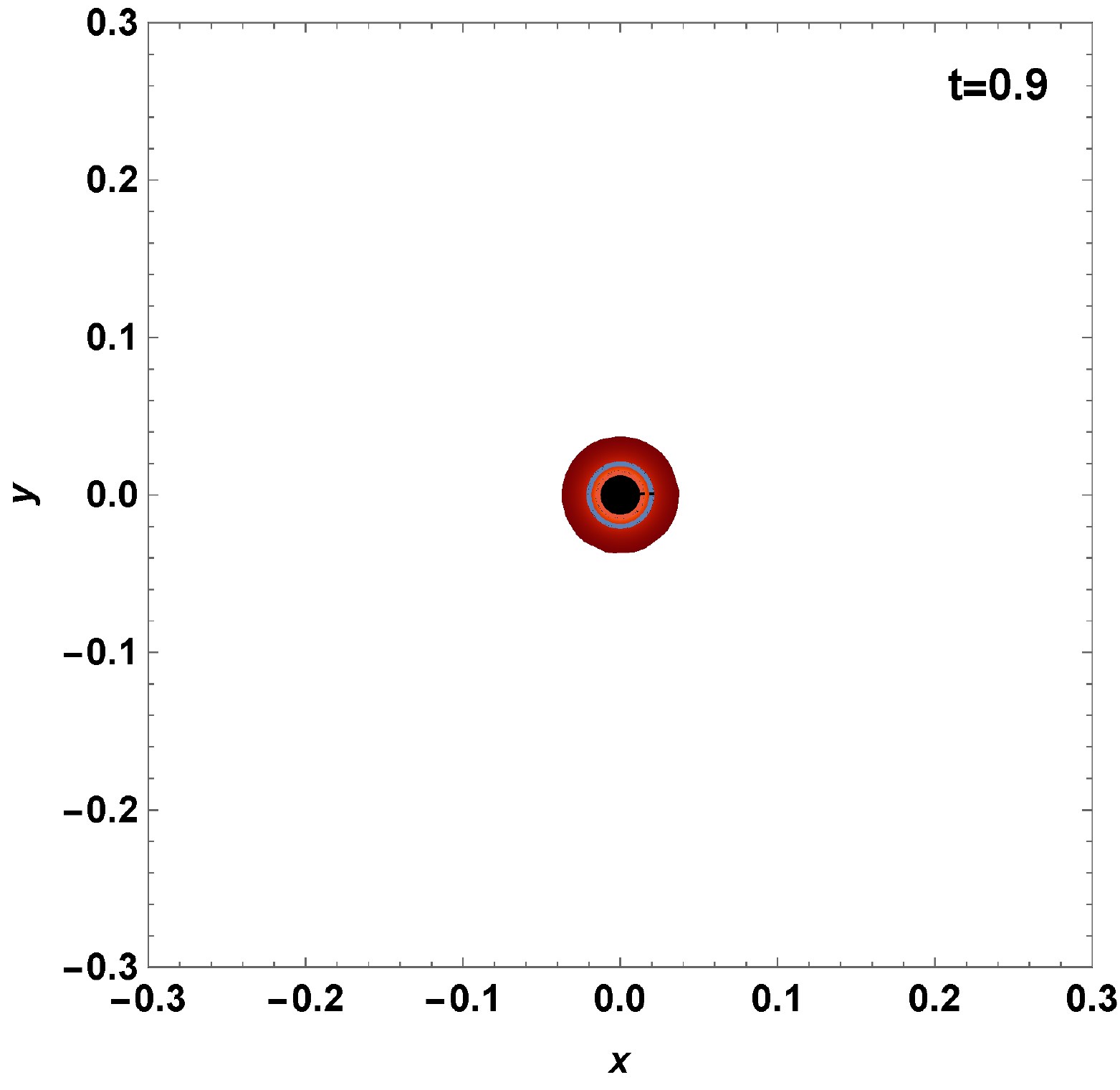}\includegraphics[scale=0.315]{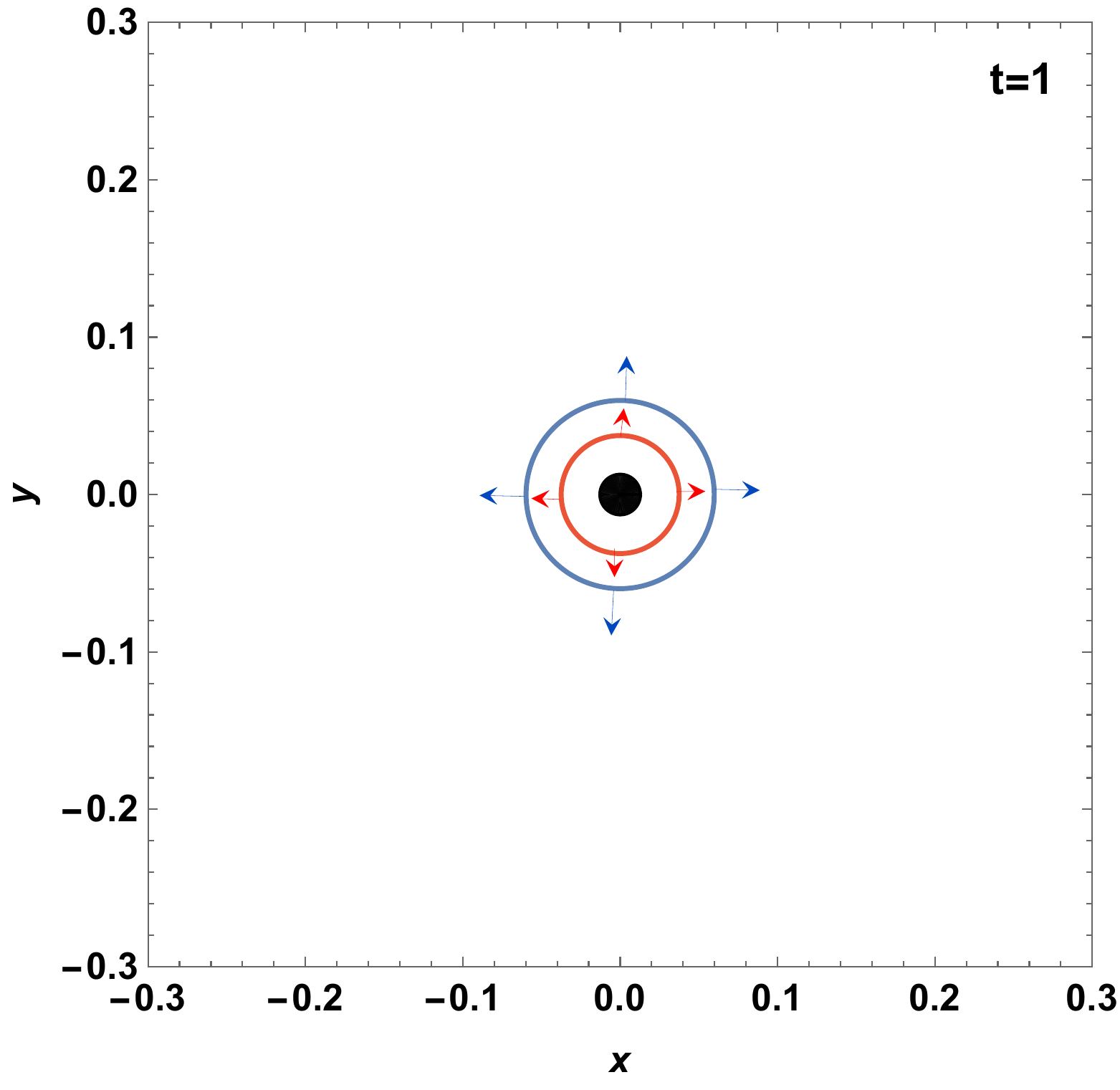}}
\caption{Evolution of the collapsing star and the global causal structure is depicted here. $F_0=1$, $F_3=-20$ and $F_i=0$ for $i \neq 1,3$. $b_{00}=-0.001$ and $b_{0j}=0$ for $j \neq 0$. $\frac{fR}{F}\sim10^{-3}$ initially and reduces in magnitude thereafter. Higher-order  terms: $o(y_1^3)$ and $o(y_1^2)$, arising in Eq.(\ref{R(t,r)}) are neglected.} The singularity is Tipler strong with $\chi_1=\chi_2=0$ and $\chi_3\neq 0$. The solid black disk represents the event horizon which increases in size with time. Escaping singular null geodesic wave fronts are represented by red and blue circles which increases with time.
\end{figure*}
However, for $\alpha=3$, if at all $\chi_1$ or $\chi_2$ is/are nonzero, then $X_0$ blows up. Hence, we will have to make sure that $\chi_1$ and $\chi_2$ should be zero. More specifically, $\chi_1$ and $\chi_2$ should be of order at least $r^3$ and $r^2$, respectively, to avoid blowing up of $X_0$ . The integral expression of $\chi_1$, $\chi_2$ and $\chi_3$ is as follows 
\cite{joshi2, mosani}:
\begin{equation}\label{chi1}
    \chi_1(v)=-\frac{1}{2}\int^{1}_{v} \frac{\frac{M_1}{v}+b_{01}}{\left(\frac{M_0}{v}+b_{00}\right)^{\frac{3}{2}}}dv, 
\end{equation}
\begin{equation}\label{chi2}
    \chi_2(v)=\int^{1}_{v}\left[\frac{3}{8}\frac{\left(\frac{M_1}{v}+b_{01}\right)^2}{\left(\frac{M_0}{v}+b_{00}\right)^{\frac{5}{2}}}-\frac{1}{2}\frac{\frac{M_2}{v}+b_{02}}{\left(\frac{M_0}{v}+b_{00}\right)^{\frac{3}{2}}}\right] dv
\end{equation}
and
\begin{widetext}
\begin{equation}\label{chi3}
     \chi_3=  \int_v^1 \frac{b_{01}}{\left(\frac{M_0}{v}+b_{00}\right)^{\frac{3}{2}}}\left(-\frac{5}{16}\left(\frac{b_{01}}{\frac{M_0}{v}+b_{00}}\right)^2 + \frac{3}{4}\left(\frac{\frac{M_2}{v}+b_{02}}{\frac{M_0}{v}+b_{00}}\right)\right)-\frac{1}{2}\frac{\left(\frac{M_3}{v}+b_{03}\right)}{\left(\frac{M_0}{v}+b_{00}\right)^{\frac{3}{2}}}dv.
\end{equation}
\end{widetext}

\begin{table}[t]
\begin{tabular}{|l|c|c|c|c|c|}
\hline
$b_{00}$  & $F_3$&  $t_{EH}(0)$\\
\hline
\hline
    $10^{-1}$     & -1  & 0.146174    \\
\hline
   $10^{-1}$       & -5   & 0.586922    \\
\hline
   $10^{-1}$    & -20  & 0.646240   \\
\hline 
   $10^{-1}$     & -50  &  0.646667  \\
\hline
 $10^{-1}$      & -100  & 0.646667 \\
\hline 
 $10^{-1}$      & -200  & 0.646667   \\
\hline 
 $10^{-2}$     & -1  & 0.170522    \\
\hline
 $10^{-2}$      & -5   & 0.609225   \\
\hline
  $10^{-2}$       & -20  & 0.664501   \\
\hline 
  $10^{-2}$       & -50  & 0.664667   \\
\hline
 $10^{-2}$      & -100  & 0.664667   \\
\hline 
  $10^{-2}$       & -200  & 0.664668    \\
\hline 
\end{tabular}
\quad
\begin{tabular}{|l|c|c|c|c|c|}
\hline
$b_{00}$  & $F_3$&  $t_{EH}(0)$\\
\hline
\hline
 $10^{-3}$       & -1  & 0.172916    \\
\hline
$10^{-3}$      & -5   & 0.611443   \\
\hline
$10^{-3}$     & -20  & 0.666319   \\
\hline 
$10^{-3}$     & -50  & 0.666467   \\
\hline
$10^{-3}$     & -100  & 0.666467   \\
\hline 
$10^{-3}$  & -200  & 0.666468   \\
\hline 
$10^{-4}$     & -1  & 0.173141   \\
\hline
$10^{-4}$     & -5   & 0.611663   \\
\hline
$10^{-4}$     & -20  & 0.666500   \\
\hline 
$10^{-4}$     & -50  & 0.666647   \\
\hline
$10^{-4}$     & -100  & 0.666648    \\
\hline 
$10^{-4}$     & -200  & 0.666647    \\
\hline 
\end{tabular}
 \caption{Here, mass function and velocity function of Eq.(\ref{mfvf}) are considered. $F_0=1$ and $t_s(0)=\frac{2}{3}$. The collapse is unbound (hyperbolic). The singularity thus formed is strong and globally hidden since $t_{EH}(0)<\frac{2}{3}$. $t_{EH}(0)$ is achieved by numerical approximation rounded up to six decimal digits.}
\label{tab1}
\end{table}

Here $M_i$ are the components nonminimally coupled to $r^i$ in the Taylor expansion of $M$ around $r=0$. $M$ is the mass profile, having relation with the Misner-Sharp mass function, as shown in Eq.(\ref{regularitycondition}). Also $b_{0i}$ in Eqs.(\ref{chi1})-(\ref{chi3}) are the components nonminimally coupled with $r^{i}$ in the Taylor expansion of the velocity profile $b_0(r)$ around the center $r=0$. Regularity condition dictates that $f(r)=r^2b_0(r)$ 
The mass profile and the velocity profile together determine the polarity of $\chi_3$. For positive $\chi_3$, we have a strong at least locally naked singularity provided $\chi_1$ and $\chi_2$ vanish at $v=0$. Such analysis was not done in \cite{deshingkar2} in the case of marginally bound collapse for various mass functions considered therein. One such example of mass function and velocity function for which $\chi_1$ and $\chi_2$ vanish is given as follows:
\begin{equation}\label{mfvf}
    F=F_0 r^3+F_3 r^6, \hspace{1cm} f=b_{00}r^2.
\end{equation}
The boundary of the cloud is found such that the density smoothly matches to zero there. Hence, the boundary is given by
\begin{equation}
    r_c=\left(-\frac{F_0}{2F_3}\right)^{\frac{1}{3}}.
\end{equation}
Similar to the previous mass function, this mass function, along with a positive velocity, also gives at least a locally naked singularity for chosen values of $F_0$ and $F_3$. $\chi_3>0$ in this case. However, in Figure 2(a), outgoing singular radial null geodesics having positive tangent at the center later gets trapped and falls back to the singularity. Increasing the magnitude of the inhomogeneity term, $F_3$, alters the evolution of the event horizon in such a way that its initiation now coincides with the time of formation of singularity due to collapsing central shell, thereby allowing singular null geodesics to escape and reach the faraway observer, as observed in Figure 2(b). 

In the case of a marginally bound collapse of dust, third-order inhomogeneity in the mass profile can give globally naked singularity for a wide range of $F_3<0$ \cite{deshingkar}. It can be seen from Eqs.(\ref{chi1})-(\ref{chi3}) that such singularity is Tipler strong.

In the case of unbound collapse, we consider a velocity function to have a positive value. It is found in Figure (3) that the mass function giving rise to the globally naked singularity as the end state of bound collapse gives a globally hidden singularity as the end state of unbound collapse having velocity function with the same magnitude but opposite polarity. Furthermore, it is observed in Table. \ref{tab1} that at least so long as the mass function and the velocity function are of the form Eq.(\ref{mfvf}) along with $b_{00}>0$, a wide range of coefficients in such mass and velocity function give a globally hidden singularity as the end state.

In Figure 4, dynamics of the collapse of the fluid are shown for a particular mass function such that the outgoing radial null geodesics get trapped, and there is no causal connection between the singular region and the outside observer. The singularity thus obtained is, however, locally naked, as seen in Figure (5). Figure (6) depicts the evolution of the density profile, event horizon, and singular geodesics escaping the boundary of the cloud without getting trapped by any trapped surfaces. A different value of the mass function is considered here. The magnitude of the inhomogeneity term in the Misner-Sharp mass function is more in this case.  An asymptotic observer may observe the wave fronts of the escaped singular null geodesic highly redshifted. Null geodesic escaping from closer to the singularity will be more redshifted. The light traveling from more close to the singularity is also traveling closer to the event horizon. One could deduce that more redshifted the light is, more significant it is, in respect of holding traces of the quantum gravity. All the evolutions are in the comoving frame. Figures 4-6 help in visualizing the evolution of the collapsing cloud along with the evolution of the event horizon and null trajectories. They also depict the dynamics of density variation of the collapsing cloud due to inhomogeneous mass distribution, bright light indicating denser.
\section{Concluding Remarks}
Some concluding remarks and open concerns are mentioned as follows:
\begin{enumerate}
    \item End state of a marginally bound collapse has been studied in \cite{deshingkar}. Considering $f=0$ eases the integration of Eq.(\ref{efe2}) to obtain the expression of $R$ as in Eq.(\ref{R(t,r)mb}).  However, such a scenario is a very particular case  corresponding to a very specific dynamics of the collapse, as mentioned in the Introduction, with a scaling function expressed as 
    \begin{equation}
        v(t,r)=\left(1-\frac{3}{2}\frac{\sqrt{F}t}{r^{\frac{3}{2}}}\right)^{\frac{2}{3}},
    \end{equation}
    which is obtained from Eq.(\ref{R(t,r)mb}). Here we consider a nonmarginally bound collapse of the inhomogeneous dust cloud and study the causal structure of the singularity formed as the end state. Investigating the nonmarginally bound gravitational collapse increases our scope of understanding the gravitational collapse to a more general scenario. It is worth mentioning that such a general scenario also encapsulates a very important case wherein, initially, all the fluid elements are at rest, i.e., $\dot R(0,r)=0$. This is obtained by considering the velocity function as 
    \begin{equation}
        f=-\frac{F}{r}
    \end{equation}
    which is obtained by substituting $\dot R(0,r)=0$ in Eq.(\ref{efe2}). Such momentarily static initial condition is motivated from the idea that collapse to a singularity begins when some dynamical instability sets in, as discussed in \cite{miyamoto, brown}.
    
    \item Unless the globally naked singularity is physically strong, it should not be taken as a serious counterexample to the weak cosmic censorship. It is important to note that the strength of the singularity as defined by Tipler 
    \cite{tipler} 
    involves vanishing of the volume element formed by three independent Jacobi fields along the timelike geodesic as it terminates in a strong singularity, rather than the behavior of individual Jacobi fields, as pointed out by Nolan 
    \cite{nolan}.
    One can show examples of physically strong singularity wherein the volume element does not vanish and hence are classified as ``Tipler weak". This led Ori 
    \cite{ori} 
    to redefine the physically strong singularity which extends the class of strong singularity by including cases in which any of the Jacobi fields is unbounded.  
    \cite{nolan, ori}.
    Such singularities are termed as ``deformationally strong" singularity. However, here we have taken an interest in singularities wherein the volume element defined by independent Jacobi fields vanishes (Tipler strong). We have proved the existence of such Tipler strong singularities that are globally naked and formed due to bound gravitational collapse.
    
    \item In deriving the explicit expression of the physical radius in terms of $t$ and $r$ in Eq.(\ref{R(t,r)}) for nonvanishing velocity function, only the first component of the Taylor expansion of $G$ is used from Eq.(\ref{taylorexpandedG}). Hence the accuracy of our further analysis will get affected for large values of the term $\frac{fR}{F}$. To minimize the error, small values of the magnitude of the velocity function are considered. For larger values, higher-order terms in the expansion of $G$ from Eq.(\ref{taylorexpandedG}) will have to be taken into account. Once the explicit expression of the physical radius is achieved, one can study the dynamics of the event horizon, apparent horizon, and singular radial null geodesics to investigate the global causal structure of the singularity.

   \item It is the event horizon, which evolves like an outgoing radial null geodesic, which starts from the singularity satisfying the equality of the physical radius and the Misner-Sharp mass function at the boundary of the collapsing fluid. Hence, any outgoing radial null geodesic with the property that $F<R$ at $r=r_c$ has to start from the center at a time before the formation of the singularity. However, this time difference between the escape of the light and the formation of the singularity can be reduced as much as desired. For such a null geodesic to be singular, it should escape from the region, which is in a small neighborhood of the singularity. This small neighborhood should have a measure of the order of Planck length. Only then will such untrapped null geodesic be considered significant and will be expected to contain traces relevant to deepen our understanding of how gravity works in the quantum regime.
    
    \item In terms of observational significance,  if at all there exists a globally visible singularity, it may be difficult to distinguish between singular and nonsingular geodesics escaping such singularity and received by a telescope. However, light wave front, which is more redshifted, is expected to come from the region, which is more close to the singularity as compared to the wave front, which is less redshifted. 
    
    \item Consider Eq.(\ref{mfvf}) with negative $F_3$ and positive $b_{00}$. This corresponds to the unbound collapse of fluid with third-order inhomogeneity in mass profile. It is found that as far as such mass and velocity functions are considered,  we may have $t_{EH}(0)<t_s(0)$, which means that globally visible singularity may not be achieved. This argument is supported by data in Table \ref{tab1}. So far, no concrete statement about the global visibility of a strong singularity formed due to unbound collapse of dust can be made, and further investigation is needed. It may be possible that for some other combination of mass function and velocity function (unbound), the collapse ends in a globally visible singularity. This will be investigated in more detail in our future work.
    
    \item A very important concern is that our analysis is restricted to the end state of a collapsing dust cloud, i.e., the pressure of the collapsing fluid is considered to be zero. The effect on the global causal structure of the singularity in the presence of pressure is unknown. To understand the behavior of singular null geodesic numerically requires information about the explicit expression of the physical radius. However, this is difficult to obtain when the Misner-Sharp mass function varies with time, which is the case when there is nonzero pressure. Investigating the global visibility of a Tipler strong singularity formed due to the collapse of a cloud having such time-varying Misner-Sharp mass function will be a significant step toward understanding the cosmic censorship.  
\end{enumerate}

\section{Acknowledgement}
K.M. would like to acknowledge the support of the Council of Scientific and Industrial Research (CSIR, India, Ref: 09/919(0031)/2017-EMR-1) for funding the work.

\end{document}